\documentclass{article}

\usepackage{microtype}
\usepackage{graphicx}
\usepackage{subfigure}
\usepackage{booktabs} 
\usepackage{tablefootnote}

\usepackage{hyperref}
\hypersetup{
    colorlinks=true,
    urlcolor=blue,
    }

 \usepackage[accepted]{icml2024}

\usepackage{amsmath}
\usepackage{amssymb}
\usepackage{mathtools}
\usepackage{amsthm}

\usepackage[capitalize,noabbrev]{cleveref}

\usepackage{natbib}

\usepackage{enumitem}

\usepackage{CJKutf8}

\usepackage{caption}
\setlist[itemize]{noitemsep}

\theoremstyle{plain}

\theoremstyle{definition}

\theoremstyle{remark}

\icmltitlerunning{Trends in AI Supercomputers}

\begin{document}
\onecolumn

\icmltitle{Trends in AI Supercomputers}

\begin{icmlauthorlist}
\icmlauthor{Konstantin F. Pilz \textbf{$^*$}}{Georgetown}
\icmlauthor{James Sanders}{Epoch}
\icmlauthor{Robi Rahman}{Epoch}
\icmlauthor{Lennart Heim}{Epoch, GovAI}
\end{icmlauthorlist}

\vskip 0.3in
\makeatletter
\renewcommand{\ICML@appearing}{}
\renewcommand{\Notice@String}{}
\makeatother

{\let\thefootnote\relax\footnotetext{%
$^1$Georgetown University
$^2$Epoch AI
$^3$Centre for the Governance of AI
\textbf{$^*$}Correspondence to Konstantin at kfp15@georgetown.edu. For inquiries about our dataset and future updates, please contact Robi at robi@epoch.ai.
}}

\begin{abstract}
\label{sec:abstract}

Frontier AI development relies on powerful AI supercomputers, yet analysis of these systems is limited. We create a dataset of 500 AI supercomputers from 2019 to 2025 and analyze key trends in performance, power needs, hardware cost, ownership, and global distribution. 
We find that the computational performance of AI supercomputers has doubled every nine months, while hardware acquisition cost and power needs both doubled every year. The leading system in March 2025, xAI's Colossus, used 200,000 AI chips, had a hardware cost of \$7B, and required 300 MW of power---as much as 250,000 households. As AI supercomputers evolved from tools for science to industrial machines, companies rapidly expanded their share of total AI supercomputer performance, while the share of governments and academia diminished. Globally, the United States accounts for about 75\% of total performance in our dataset, with China in second place at 15\%.
If the observed trends continue, the leading AI supercomputer in 2030 will achieve $2\times10^{22}$ 16-bit FLOP/s, use two million AI chips, have a hardware cost of \$200 billion, and require 9 GW of power. 
Our analysis provides visibility into the AI supercomputer landscape, allowing policymakers to assess key AI trends like resource needs, ownership, and national competitiveness.

\end{abstract}

\newpage


\section*{Executive Summary}\label{sec:executive-summary}

\textbf{AI progress has relied on exponentially larger AI supercomputers.} The compute used to train the most notable AI models has grown by 4.1$\times$ per year since 2010, enabling breakthroughs like advanced chatbots, image generation, and protein structure prediction. This training compute growth relied primarily on larger AI supercomputers that now consist of more than 100,000 AI chips, have hardware costs of billions of dollars, and consume as much power as a medium-sized city. We compile a dataset of over 500 AI supercomputers worldwide by systematically collecting public data from 2019 to 2025. We define an AI supercomputer as a system using AI chips that achieved at least 1\% of the computational performance of the leading AI supercomputer when it first became operational. We estimate our dataset captures 10-20\% of all existing AI supercomputer capacity, based on comparing the total performance to public AI chip production and sales estimates.

\textbf{The computational performance of leading AI supercomputers has doubled every 9 months, driven by deploying more and better AI chips }(Figure \ref{fig:abstract:Computational_Performance_of_Supercomputers_since_2019}). Two key factors drove this growth: a yearly 1.6$\times$ increase in chip quantity and a 1.6$\times$ annual improvement in performance per chip. While systems with more than 10,000 chips were rare in 2019, companies deployed AI supercomputers more than ten times that size in 2024, such as xAI’s Colossus with 200,000 AI chips.

\vspace{3mm}
{\setlength\intextsep{0pt}
\begin{figure}[H]
     \centering
     \includegraphics[width=1\linewidth]{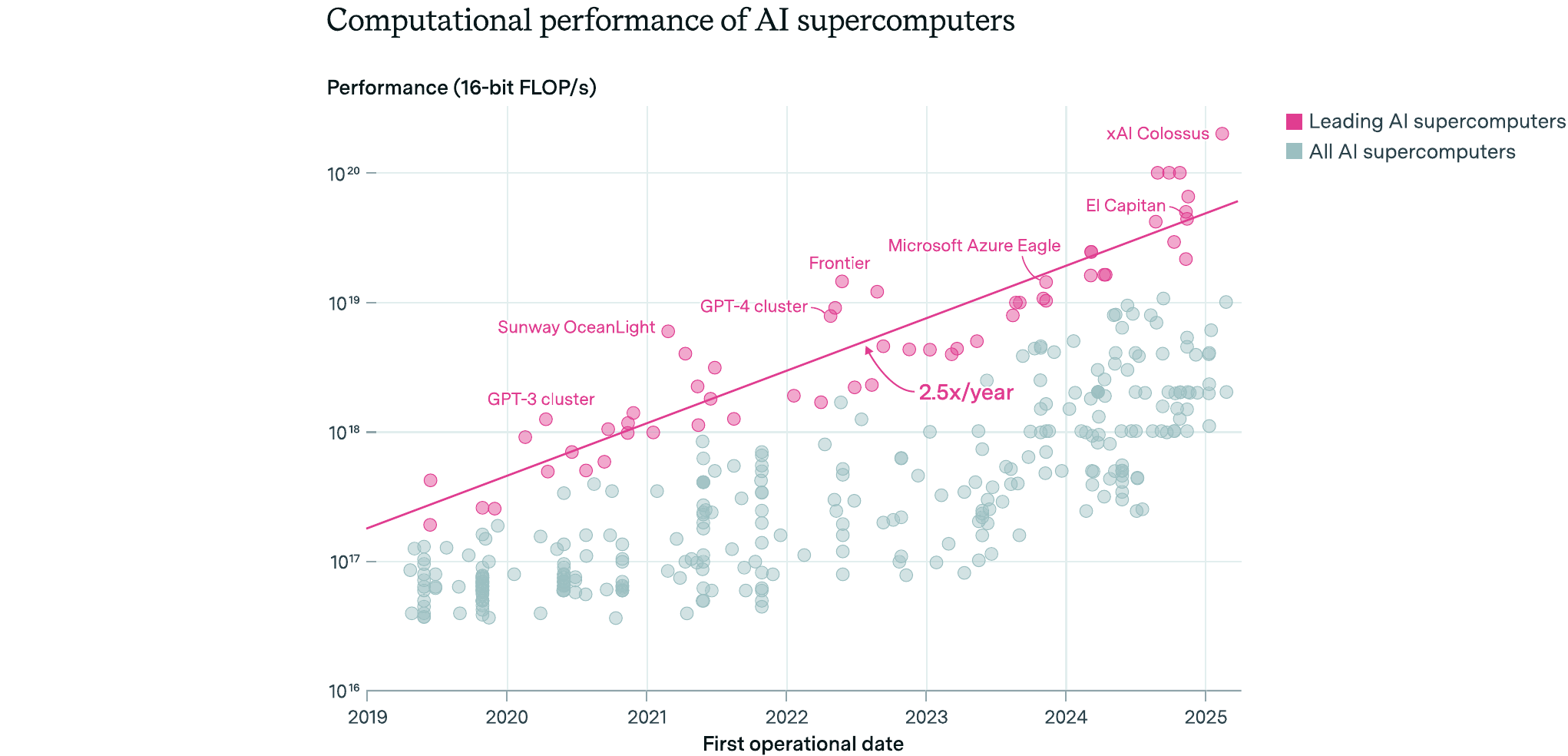}
     \caption{The performance of leading AI supercomputers (in FLOP/s, for 16-bit precision) has doubled every 9 months (a rate of 2.5$\times$ per year).}
     \label{fig:abstract:Computational_Performance_of_Supercomputers_since_2019}
\end{figure}
}

\textbf{Power requirements and hardware costs of leading AI supercomputers have doubled every year.} Hardware cost for AI supercomputers has increased by 1.9$\times$ every year, while power needs increased by 2.0$\times$ annually. As a consequence, the most performant AI supercomputer as of March 2025, xAI’s Colossus, had an estimated hardware cost of \$7 billion (Figure \ref{fig:abstract:Cost_of_supercomputers_since_2019}) and required about 300 MW of power---as much as 250,000 households. Alongside the massive increase in power needs, AI supercomputers also became more energy efficient: computational performance per watt increased by 1.34$\times$ annually, which was almost entirely due to the adoption of more energy-efficient chips.

\noindent
\textbf{If the observed trends continue, the leading AI supercomputer in June 2030 will need 2 million AI chips, have a hardware cost of \$200 billion, and require 9 GW of power.} Historical AI chip production growth and major capital commitments like the \$500 billion Project Stargate suggest the first two requirements can likely be met. However, 9 GW of power is equivalent to 9 nuclear reactors, a scale beyond any existing industrial facility. To overcome power constraints, companies may increasingly use decentralized training approaches, which would allow them to distribute a training run across AI supercomputers in several locations.

{\setlength\intextsep{0pt}
\begin{figure}[H]
     \centering
     \includegraphics[width=1\linewidth]{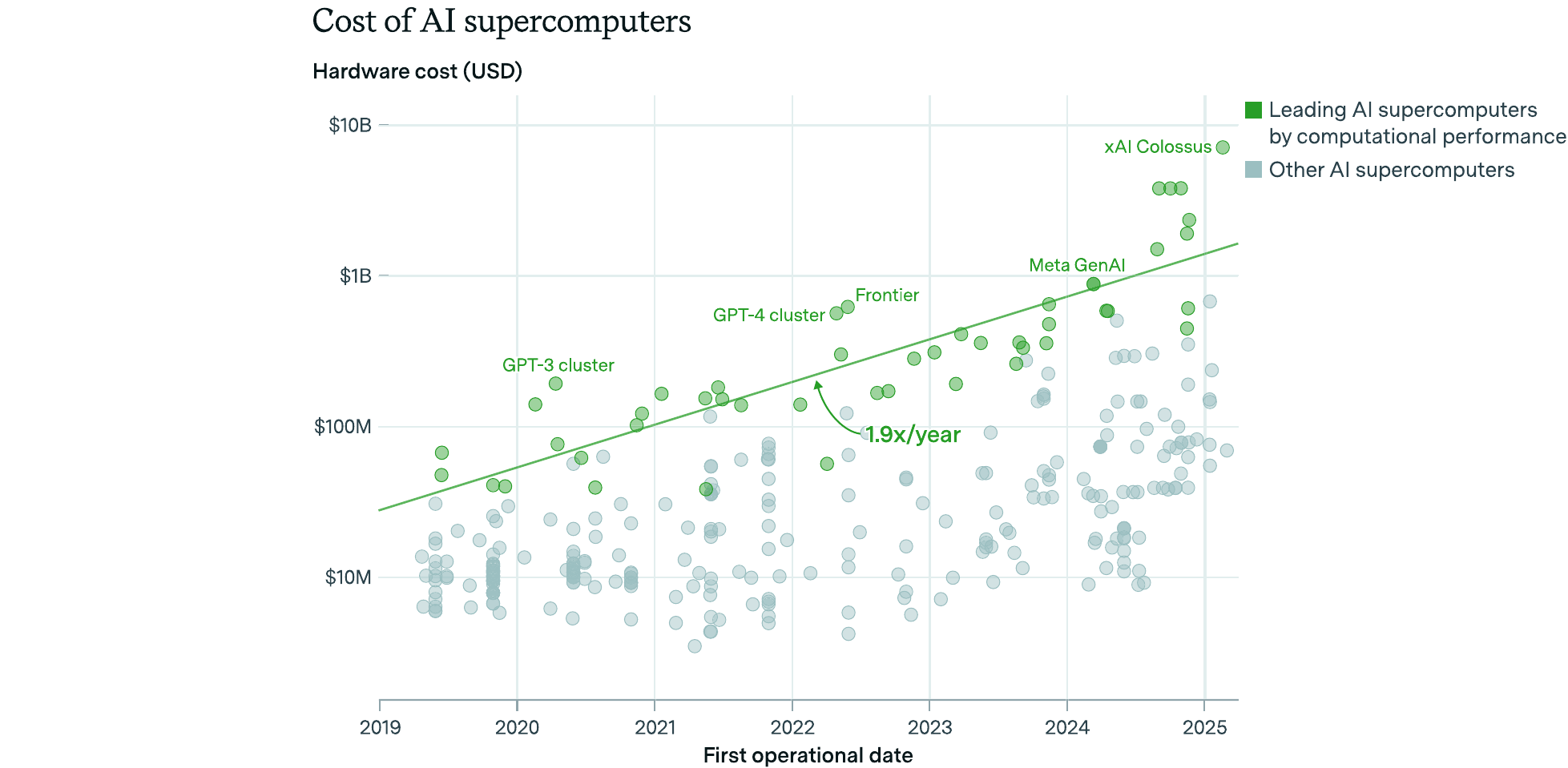}
     \caption{The cost of leading AI supercomputers (in 2025 USD) has doubled roughly every year.}
     \label{fig:abstract:Cost_of_supercomputers_since_2019}
\end{figure}
}

\textbf{Companies now dominate AI supercomputers.} As AI development has attracted billions in investment, companies have rapidly scaled their AI supercomputers to conduct larger training runs. This caused leading industry system performance to grow by 2.7$\times$ annually, much faster than the 1.9$\times$ annual growth of public sector systems. In addition to faster performance growth, companies also rapidly increased the total number of AI supercomputers they deployed to serve a rapidly expanding user base. Consequently, industry's share of total AI compute surged from 40\% in 2019 to 80\% in 2025, as the public sector’s share fell below 20\% (Figure \ref{fig:abstract:Performance_of_public_vs_private_supercomputers}).

\vspace{1mm}
{\setlength\intextsep{0pt}
\begin{figure}[H]
    \centering
    \includegraphics[width=0.85\linewidth]{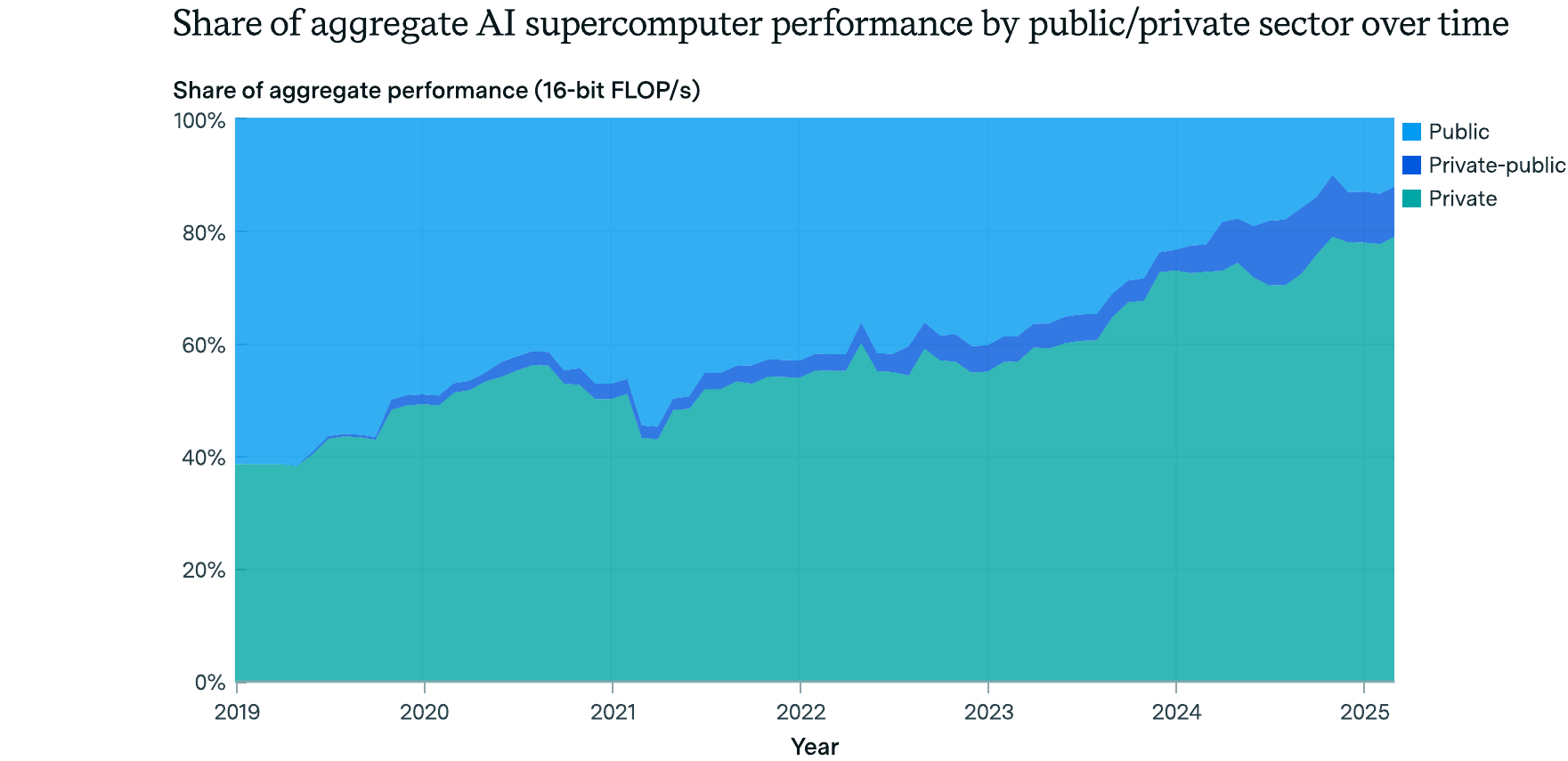}
    \caption{Share of aggregate AI supercomputer performance over time in the public vs private sector.}
    \label{fig:abstract:Performance_of_public_vs_private_supercomputers}
\end{figure}
}

\textbf{The United States hosts 75\% of AI supercomputers, followed by China.} The United States accounts for about three-quarters of total AI supercomputer performance, with China in second place at 15\% (Figure \ref{fig:abstract:Distribution_of_supercomputer_over_time_by_country}). Meanwhile, traditional supercomputing powers like the UK, Germany, and Japan now play marginal roles in AI supercomputers. This shift reflects the dominance of large, U.S.-based companies in AI development and computing. However, AI supercomputer location does not necessarily determine who uses the computational resources, given that many systems in our database are available remotely, such as via cloud services.

\vspace{3mm}
{\setlength\intextsep{0pt}
\begin{figure}[H]
    \centering
    \includegraphics[width=0.85\linewidth]{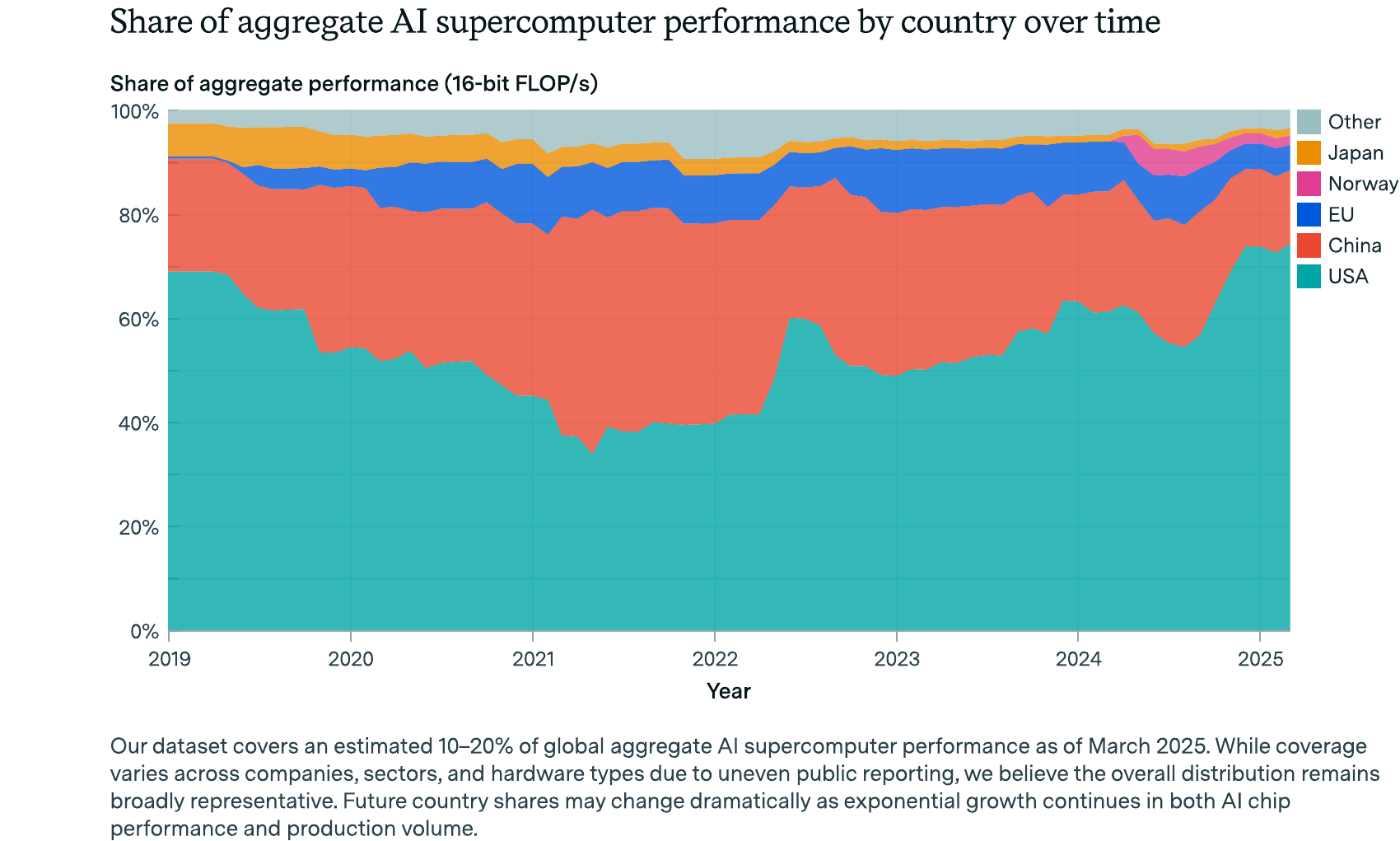}
    \caption{Share of AI supercomputer computational performance by country over time. We are visualizing all countries that held more than a 3\% share at some point in time. 
}
    \label{fig:abstract:Distribution_of_supercomputer_over_time_by_country}
\end{figure}
}

\textbf{We are releasing our }\href{https://epoch.ai/data/ai-supercomputers}{\textbf{dataset}}\textbf{ along with }\href{https://epoch.ai/data/ai-supercomputers-documentation}{\textbf{documentation}} soon after this publication. Our data will be part of Epoch AI's \href{https://epoch.ai/data}{Data on AI} hub and maintained with regular updates.



\newpage
\setcounter{tocdepth}{3}
\tableofcontents
\newpage


\section{Introduction}\label{sec:Introduction}

The computing resources (compute) used to train notable AI models have increased at a rate of 4--5$\times$ per year since the beginning of the deep learning era in 2010 \cite{sevilla2024compute}. This exponential increase has been a major driver of improvements in AI capabilities across many domains, such as in large language models or image generation \cite{erdil2022compact, ho2024capabilities}. Most of this increase in compute has been driven by larger, higher-performance AI supercomputers \citep{hobbhahn2023progress, frymire2024evolution}. 

Given their importance for AI development, systematically collecting data on AI supercomputers allows us to better understand trends such as their hardware costs, power requirements, and global distribution. This analysis is relevant to policymakers, because compute is both an enabler of AI progress and a potential tool for governance \citep{sastry2024compute, khan2020ai}. For instance, information about the distribution of AI supercomputers across countries allows governments to assess their national competitiveness in AI, and data on the growth of power requirements can help with electrical grid planning.

However, despite the importance of AI compute, no comprehensive dataset of AI-specific supercomputers exists. Resources like the Top500 list or the ML-Perf benchmark rely on voluntary submissions and thus lack sufficient data to reliably analyze trends \citep{top500undated}.\footnote{For a review of existing data sources, see Appendix \ref{sec:Appendix:Review_of_existing_data}.} Meanwhile, databases used for business intelligence, such as SemiAnalysis’s data center model, are not available for public analysis and focus on present-day systems rather than historical trends \citep{semianalysundateddatacentermodel}.

We attempt to close this gap by collecting data from various public sources and establishing a dataset of 500 AI supercomputers between 2019 and 2025. We use this to study several key trends: the growth of AI supercomputer performance, hardware costs, power consumption, and the distribution of AI supercomputing power between countries and sectors. 


\section{Methods}
\label{sec:Methods}

\subsection*{AI supercomputer definition}
\label{sec:Methods:AI_Supercomputer_definition}

We define an AI supercomputer as a computer system that can support training large-scale AI models, deployed on a contiguous campus. We use two criteria to assess whether a given system can support training large-scale AI models:
\begin{itemize}[topsep=0pt]
    \item[1.] The system contains chips that can accelerate AI workloads, such as NVIDIA’s V100, A100, H100, and GB200, Google’s TPUs, and other chips commonly used to train frontier AI models. To assess if a given chip is suitable for large-scale AI training, we use a dataset of machine learning hardware created by \citet{hobbhahn2023progress}. If a chip is not part of that dataset, we consider it an AI chip if it has the following features:
    \vspace*{1mm}
        \begin{itemize}[topsep=0pt]
            \item[{\textbullet}] Support for precisions commonly used in AI training, such as FP16 or INT8.
            \item[{\textbullet}] Compute units dedicated for matrix multiplications, such as tensor cores in NVIDIA GPUs.
            \item[{\textbullet}] High-bandwidth memory (HBM) or other memory types enabling a high memory bandwidth.
            \item[{\textbullet}] Was used to train a model in \citet{epochai2025notable}'s notable AI models dataset.
        \end{itemize}
    \vspace*{1mm}
    \item[2.] The system has a high theoretical computational performance on AI-relevant precisions.\footnote{We consider 32, 16, and 8-bit number formats as AI-relevant in our study period.} Due to the rapid pace of hardware improvements, we use a moving definition and only include systems that have at least 1\% of the performance of the most performant existing AI supercomputer at that time.\footnote{Our inclusion criteria compare the system's highest performance available in 32-, 16-, or 8-bit arithmetic formats to the highest performance rate of the leading AI supercomputer at the time. Note that we exclude systems that do not support 32-bit or lower precision formats from the analysis. See Appendix \ref{sec:Appendix:Methods:Numerical_precision} for details on our approach.}
\end{itemize}

To balance data collection effort and representativeness, we limit the scope of our data collection to about six years, from the start of 2019 to February 2025. We will maintain the dataset at \url{https://epoch.ai/data/ai-supercomputers} and integrate it with Epoch AI’s \href{https://epoch.ai/data}{Data on AI} hub.

\subsection*{Data collection}
\label{sec:Methods:Data_collection}

We use the Google Search API, existing compilations of (AI) supercomputers, and manual searches to collect a dataset of 501 leading AI supercomputers between 2019 and 2025. We also cover an additional 225 systems pre-2019 for a total of 726 AI supercomputers.\footnote{Our dataset includes an additional 99 systems that we exclude because they are below our inclusion threshold or otherwise outside our definition. When including these excluded systems, our total is 825 entries.} Our most significant sources are company announcements, Top500 entries with significant GPU numbers, and the \citet{epochai2025notable} dataset of notable AI models. For each potential AI supercomputer, we manually search for details such as the number and type of chips used by the system, when it was first operational, its reported performance, owner, and location.

We estimate our dataset covers about 10\% of the aggregate performance of all AI chips produced until 2025 and about 15\% of the AI chip stocks of the largest companies as of early 2025. Our dataset covers about half of all systems used in the 25 largest training runs in Epoch AI’s notable models dataset as of March 2025 \citep{epochai2025notable}. For a detailed analysis of our coverage, see Appendix \ref{sec:Appendix:Limitations:Summary}.

\subsection*{Analysis}
\label{sec:Methods:Analysis}
We combine our collected data with Epoch AI's data on machine learning hardware to estimate the total performance\footnote{We define computational performance for an AI supercomputer as the advertised theoretical maximum non-sparse FLOP/s in a given numerical precision for an AI chip, summed over all AI chips in the system.}, hardware cost, and power requirements of systems in our database \citep{EpochMachineLearningHardware2024,hobbhahn2023progress}. We filter our dataset to 389 high-certainty, confirmed operational systems between 2019-01-01 and 2025-03-01. We then fit regressions for key metrics of the 57 AI supercomputers in our study period that were in the top-10 worldwide by 16-bit FLOP/s when they first became operational. The metrics we analyze include computational performance, number of chips, power requirements, energy efficiency, and hardware costs. We further assess the distribution across sectors and countries of the aggregate performance of all AI supercomputers in our dataset, including pre-2019 systems, for a total of 470 systems. Appendix \ref{sec:Appendix:Methods} contains detailed information on our data collection, estimations for hardware cost and power, and methods for data analysis.

\section{Results}
\label{sec:Results}
We first assess growth in performance, power, and hardware cost for the leading AI supercomputers in our dataset. We then examine how AI supercomputers in our dataset are distributed across the private vs public sector, and across different countries. 

\subsection{Computational performance of the leading AI supercomputers has doubled every nine months}
\label{sec:Results:Computational_performance}
The computational performance of leading AI supercomputers increased by 2.5$\times$ per year between 2019 and 2025 (Figure \ref{fig:Computational_Performance_of_Supercomputers_scatterplot_since_2017}).\footnote{In our study period AI training workloads shifted from 32-bit precision, to 16-bit precision and partially to 8-bit precision (see Appendix \ref{sec:Appendix:Methods:Numerical_precision} for an explanation of different precisions and how we handle them). We provide an overview table of all metrics for each precision in Appendix \ref{sec:Appendix:Additional_data:Different_precisions}.} Performance increased at an even faster rate when considering only AI supercomputers owned by companies (Section \ref{sec:Results:Computational_performance:Industry_outpaced_public_sector}). The rapid increase in performance resulted in the leading system in March 2025, xAI's Colossus achieving over 50 times the performance of Oak Ridge National Laboratory's Summit, the leading AI supercomputer in 2019.\footnote{Summit achieved a performance of $3.5\times10^{19}$ FLOP/s (16-bit precision) while Colossus achieved a performance of $2.0\times10^{20}$ FLOP/s.} 

We found several large AI supercomputers in 2017 and 2018, significantly above the trend suggested by our post-2018 results. It is unclear to what extent this reflects a lack of coverage in our dataset or whether these genuinely were the largest deployed systems until 2021. We discuss in Section \ref{sec:Discussion:Supercomputer_growth_relied_on_and_enabled_AI_development} how these early systems were primarily used for scientific research, rather than for conducting large training runs, and may not be directly comparable to later systems.

{\setlength\intextsep{0pt}
\begin{figure}[H]
    \centering
    \includegraphics[width=1\linewidth]{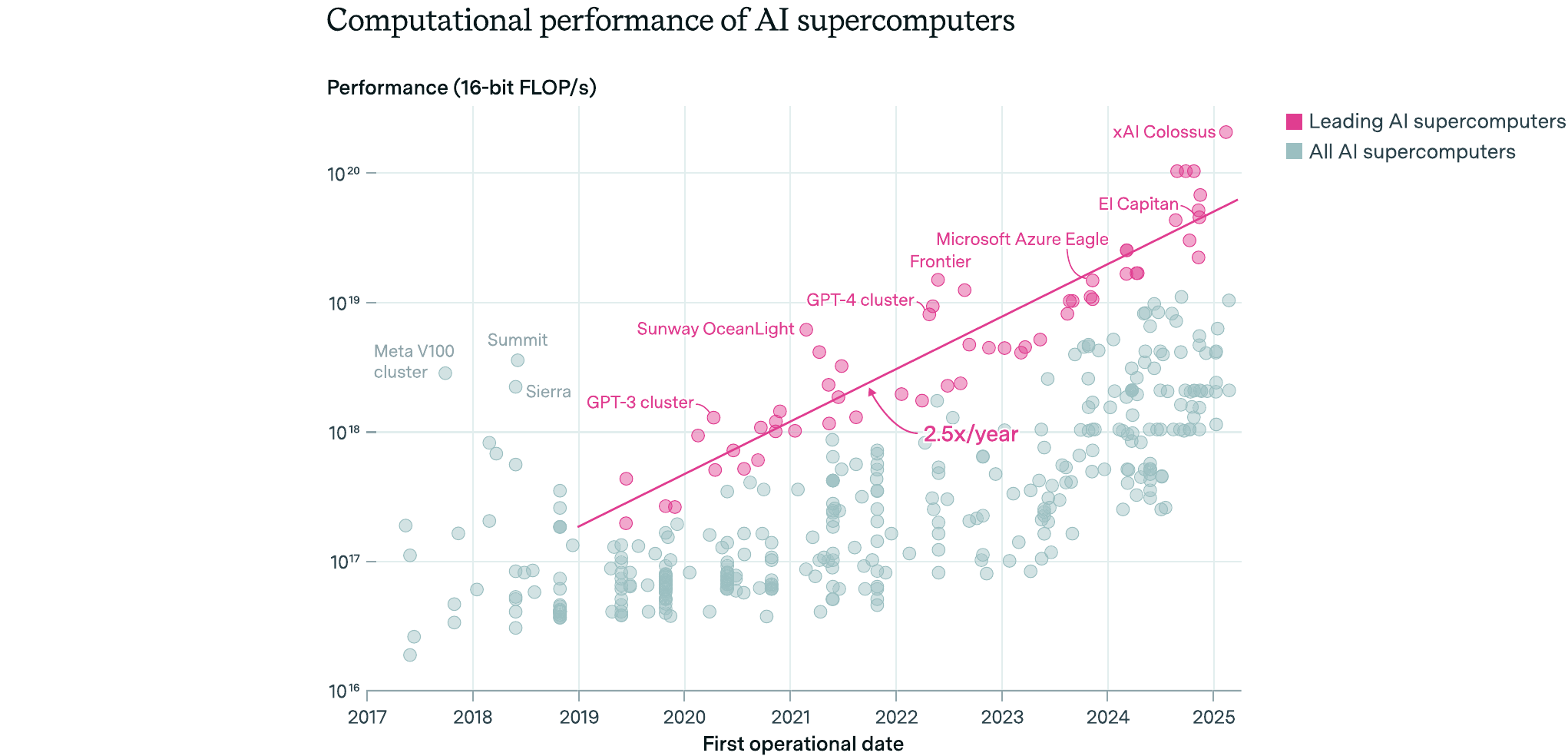}
    \caption{Performance of top-10 leading AI supercomputers, with performance measured in FLOP/s at 16-bit precision, has increased by 2.5$\times$ per year (90\% Confidence Interval (CI): 2.4--2.7$\times$). We start the regression in 2019, but consider pre-2019 AI supercomputers to identify which systems were in the top-10 at the start of 2019. Our pre-2019 data is too limited to include in the regression. We highlight some notable systems.}
    \label{fig:Computational_Performance_of_Supercomputers_scatterplot_since_2017}
\end{figure}
}
\subsubsection{Performance increases relied on AI supercomputers using more and better AI chips}
\label{sec:Results:Computational_performance:Relied_on_more_and_better_chips}
The annual performance increase of 2.5$\times$ resulted from two roughly equal factors: increased number of AI chips, and improved performance per chip.

First, the number of chips in the highest performing AI supercomputers increased by 1.6$\times$ annually (Figure \ref{fig:AI_chip_quantity_scatterplot_since_2017}). In January 2019, Oak Ridge’s Summit had the highest chip count, with 27,648 NVIDIA V100s.\footnote{Summit was the system with the highest chip count within the top-10 leading AI supercomputers by performance. Tianhe-2, a Chinese system from 2013 had a higher AI chip count of 48,000, but it was not in the top-10 AI supercomputers by performance as of January 2019.} By March 2025, xAI’s Colossus had the highest chip count of all known systems with 200,000 NVIDIA H100s and H200s.\footnote{Colossus phase 2 likely used 150k H100s and 50k H200s \citep{shilov2024colossus}.} Including pre-2019 systems in the regression would probably result in a lower growth rate. However, we cannot reliably do so because our data collection only goes back to 2019.

Second, the computational performance per chip in the most performant AI supercomputers increased by 1.6$\times$ annually. Three chip generations are notable in our study period. Between 2019 and 2021, NVIDIA’s V100 was the most prominent chip, making up more than 90\% of installed performance. In 2021, NVIDIA’s A100 gained prominence and became the most prevalent chip by 2023, with AMD’s MI250X and Google’s TPU v4 making up minority shares.\footnote{Our coverage of TPUs is limited given Google exclusively uses them internally and hardly publicizes their deployment.} In 2023, NVIDIA’s H100 became more widespread, exceeding 50\% of total performance in our dataset by July 2024.
    
The 1.6$\times$ (90\% CI: 1.5--1.7) improvement in computational performance per chip of leading AI supercomputers is slightly faster than the general trend of AI chip performance improving 1.28$\times$ per year (90\% CI: 1.24--1.32) for FP32 and 1.38$\times$ per year (90\% CI: 1.28--1.48) for FP16 \citep{rahman2025computational, hobbhahn2023progress}. This difference likely stems from AI supercomputers primarily incorporating leading AI chips rather than average-performing ones.
\vspace*{3mm}
{\setlength\intextsep{0pt}
\begin{figure}[H]
    \centering
    \includegraphics[width=1\linewidth]{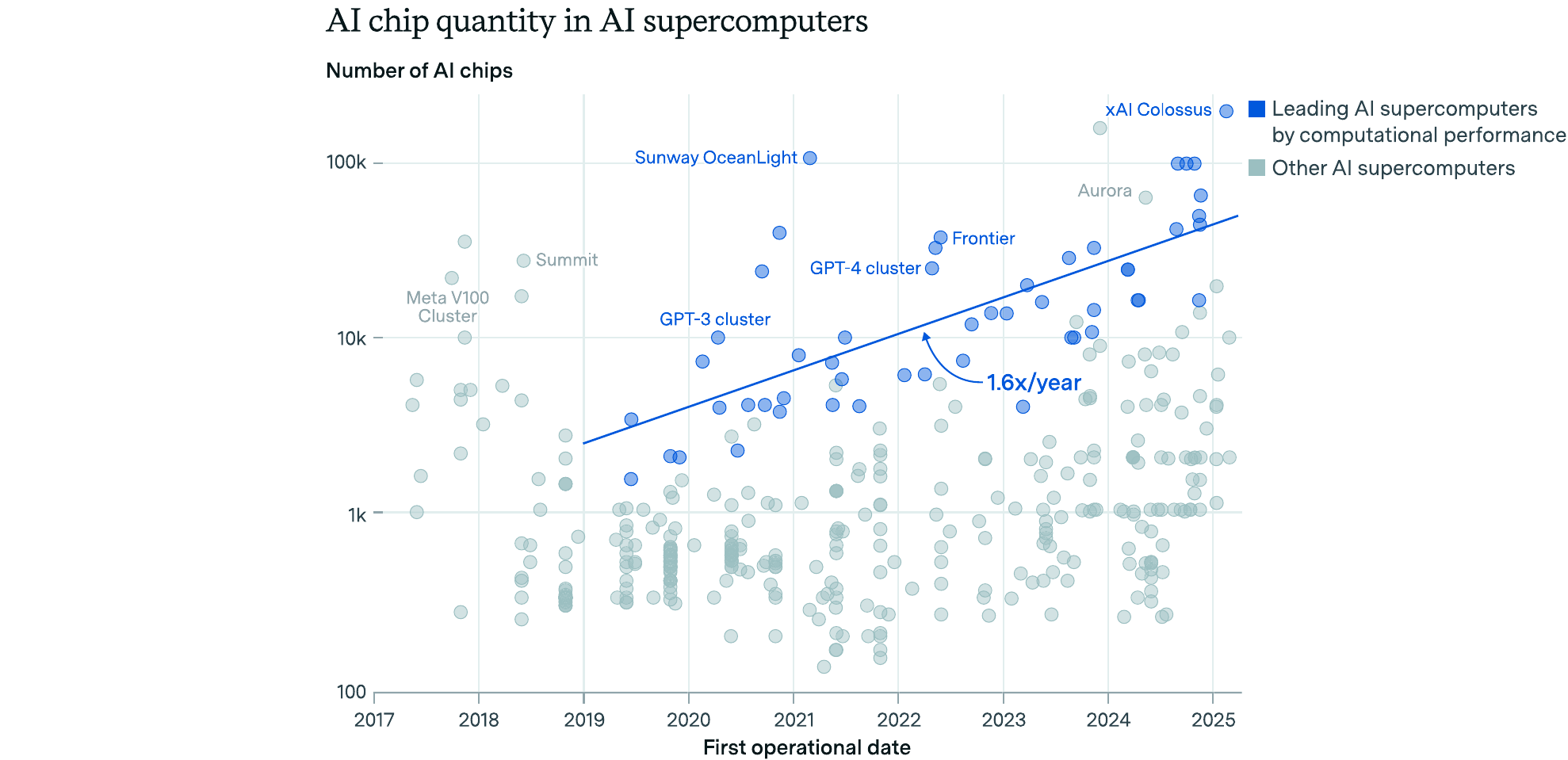}
    \caption{The number of AI chips in the leading AI supercomputers grew by 1.6$\times$ per year (90\% CI: 1.5--1.8$\times$). We start the regression in 2019, but gathered data further back to determine which 2019 AI supercomputers were in the top-10. Our pre-2019 data is too limited to include in the regression. See Section \ref{sec:Methods:Analysis} for full methods.\protect\footnotemark}
    \label{fig:AI_chip_quantity_scatterplot_since_2017}
\end{figure}
}
\footnotetext{Note on Sunway OceanLight: This Chinese system is the first AI supercomputer to cross the 100,000 AI chip threshold. However, limited details are available, and we were unsure about whether to count the processors used as AI chips because they primarily serve as CPUs. Yet the chips support 16-bit operations and the system is part of the top-10 most performant AI supercomputers of its time \citep{shilov2023}. We thus decided to include it.}
\subsubsection{AI supercomputer performance increased faster than traditional supercomputers}
\label{sec:Results:Computational_performance:AI_supercomputers_faster_than_traditional_supercomputers}

\citet{benhari2024survey} found that the 64‑bit performance of the largest Top500 supercomputer increased by 1.45$\times$ per year between 1994 and 2023.\footnote{They find that the doubling time is 1.87 years, which means performance increases by 2.35$\times$ annually: $2^{(1/1.87)}=1.45$.} This growth rate makes top‑10 AI supercomputers’ performance increase significantly faster than the historic trend for the Top500’s top machine. Two factors likely drive this divergence: AI-specific chips and faster investment growth.

First, AI chip performance has outpaced that of CPUs \citep{hobbhahn2023progress}. This is because AI computing workloads have different properties than traditional computing, allowing AI chip designers to optimize performance for parallel matrix operations, which has led to AI chip performance advancing significantly faster than CPU performance \citep{hobbhahn2023progress}.

Second, investment in AI supercomputers has increased more rapidly than investment in traditional supercomputers. The Top500 list was historically shaped by government-funded projects, which only slowly increased in budgets. However, our AI supercomputer dataset primarily captures systems owned by large companies, which have rapidly increased investment in AI supercomputers in the 2020s \citep{cottier2024historical}.

\subsubsection{AI supercomputers in private industry have outpaced those in government or academia}
\label{sec:Results:Computational_performance:Industry_outpaced_public_sector}

The performance of the leading AI supercomputers from companies grew by 2.7$\times$ annually between 2019 and March 2025. Meanwhile, the performance of the leading AI supercomputers owned and funded by governments and academic institutions grew, significantly slower, by only 1.9$\times$ annually (p = 0.022). 
The largest known public AI supercomputer, Lawrence Livermore’s El Capitan, now only achieves 22\% of the computational performance of the largest known industry AI supercomputer, xAI’s Colossus. We discuss this shift from the public to the private sector in Section \ref{sec:Discussion:Consequences_of_private_sector_dominance}

{\setlength\intextsep{0pt}
\begin{figure}[H]
    \centering
    \includegraphics[width=1\linewidth]{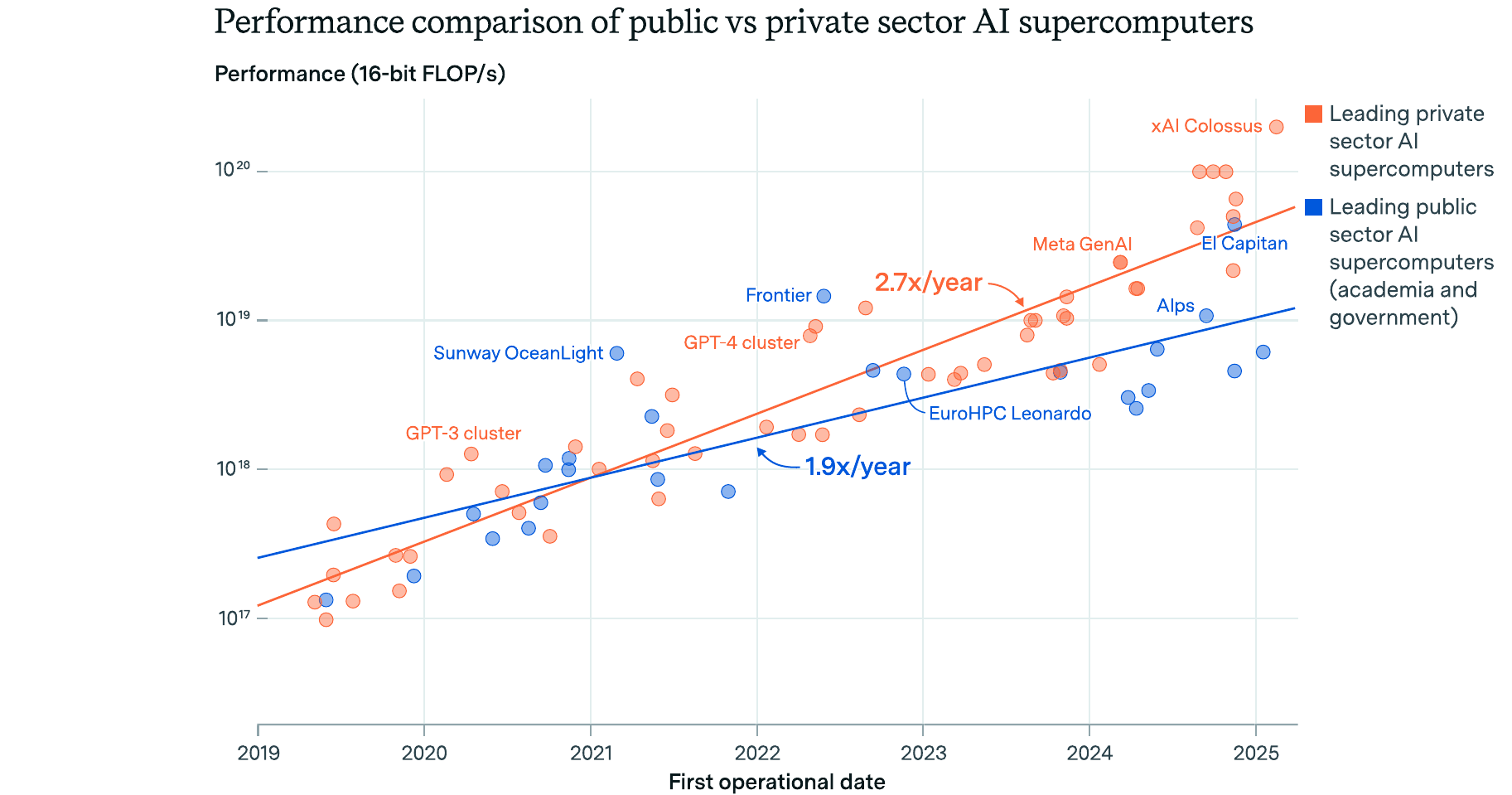}
    \caption{Performance of leading AI supercomputers owned by the private versus the public sector (government and academia). Leading public sector systems started out larger but have not kept pace with industry systems, which have grown at 2.7$\times$ annually (90\% CI: 2.5--2.9$\times$), while public sector systems have only grown at 1.9$\times$ annually (90\% CI: 1.6--2.2$\times$). Note that we exclude AI supercomputers funded and owned by both public and private institutions.}
    \label{fig:public_vs_private_supercomputers_trend_since_2019}
\end{figure}
}

\subsubsection{AI supercomputers have kept pace with 4--5$\times$ annual growth in the largest training runs} 
\label{sec:Results:Computational_performance:Supercomputer_performance_vs_training_run_required_performance}

\citet{sevilla2024compute} found that the training compute for the largest AI models has grown by 4.2$\times$ per year (90\% CI: 3.6--4.9$\times$) between 2018 and 2024. This aligns with our observed AI supercomputer performance growth, after we account for increasing training durations.\footnote{Training durations of the top-10 largest AI training runs increased by 1.4$\times$ annually between 2019 and 2025 \citep{frymire2024evolution}.}

In Figure \ref{fig:largest_supercomputers_vs_required_supercomputers_for_largest_models}, we show the required computational performance for the largest AI training runs, and the performance of leading AI supercomputers in our dataset.\footnote{Epoch's notable AI model database reports training compute (in FLOP) independent of the precision. We thus assess the performance trend considering the highest performance across 32, 16, and 8-bit, which were the most commonly used precisions for AI training between 2019 and 2025.} We consider only industry systems, which ran the vast majority of AI training runs \citep{besiroglu2024chinllm}. To calculate the performance needed for training runs, we divide training compute in FLOP by the training duration in seconds, adjusted by an average performance utilization of 40\% \citep{sevilla2022}.

Between 2019 and 2025, the largest industry AI supercomputers consistently achieved 10$\times$ the computational performance required for the largest AI training runs (not including compute required for experiments before the final training run). While the systems required for the largest training runs have grown slightly faster than the leading AI supercomputers (3.4$\times$ vs 3.0$\times$), we find no statistically significant difference in the two trends ($p=0.18$). Hence, AI supercomputer growth has been consistent with the increase in training compute, as shown in Figure \ref{fig:Sankey_diagram_drivers_of_increased_compute}.

\vspace{2mm}
{\setlength\intextsep{0pt}
\begin{figure}[H]
    \centering
    \includegraphics[width=1\linewidth]{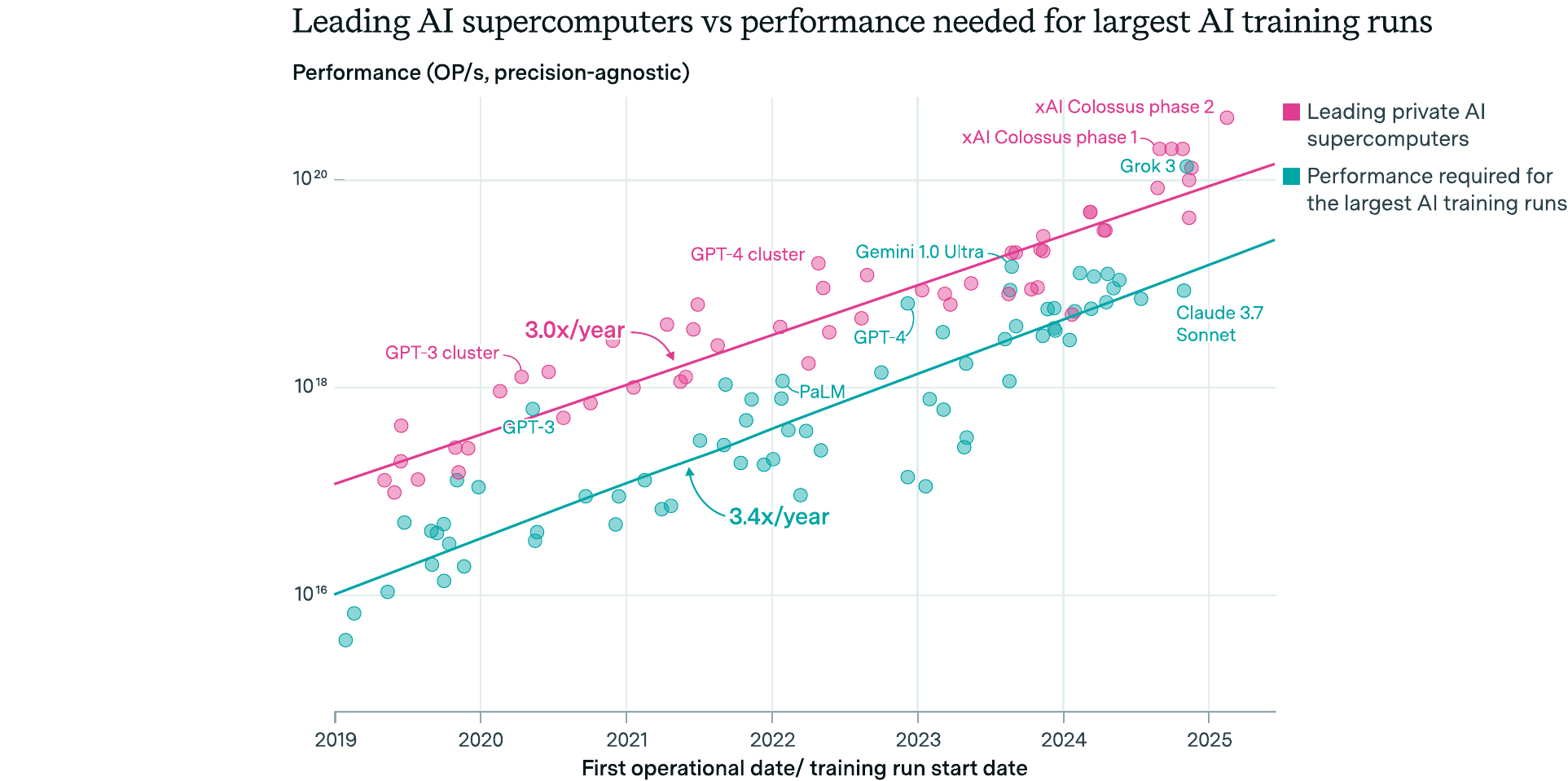}
    \caption{Computational performance of largest industry AI supercomputers and the performance required for the largest reported AI training runs \citep{epochai2025notable}. To estimate the size of the AI supercomputer needed for these training runs, we assume a GPU utilization rate of 40\% and use the stated training run durations when available, or an estimate from a regression on training durations of the largest AI models. We subtract the training duration from the publication date for notable models to better approximate when their training started. Given that the notable model dataset does not report the numerical precision used for training, we also report precision-agnostic OP/s for AI supercomputers, considering the highest performance available across 32, 16, and 8-bit number formats.}
    \label{fig:largest_supercomputers_vs_required_supercomputers_for_largest_models}
\end{figure}
}

\vspace{3mm}
{\setlength\intextsep{0pt}
\begin{figure}[H]
    \centering
    \includegraphics[width=0.8\linewidth]{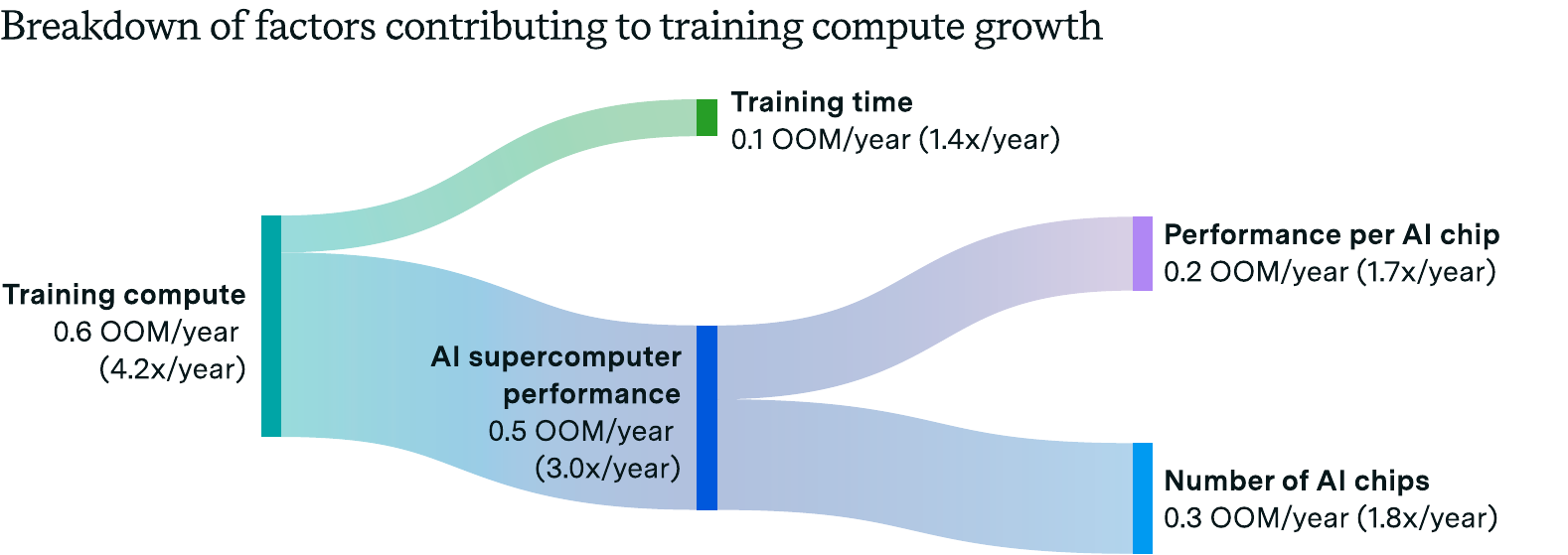}
    \caption{Overview of drivers of increasing training compute. OOM stands for orders of magnitude. AI supercomputer metrics are based on private sector systems and the highest computational performance across precisions.}
    \label{fig:Sankey_diagram_drivers_of_increased_compute}
\end{figure}
}

\subsection{Power requirements of the leading AI supercomputers doubled every 13 months}
\label{sec:Results:Power_capacity}

We assess the annual growth rate in power requirements of the leading AI supercomputers either based on reported power requirement or, if unavailable, by estimating the power requirement based on the number and type of AI chips, including additional IT infrastructure like CPUs, network switches, and data center supporting infrastructure like cooling and power conversion. For details on our power estimation, see Appendix \ref{sec:Appendix:Methods:Power_Requirements}.

We find that the power need of the leading AI supercomputers increased by 2.0$\times$ each year between 2019 and 2025. In January 2019, Summit at Oak Ridge National Lab had the highest power requirement with 13 MW.\footnote{Tianhe-2 had the highest power requirement in our dataset with 24 MW, but was not top-10 by performance.} In 2024, the first systems began to cross the 100 MW threshold and in March 2025, xAI’s Colossus had the highest power requirement at an estimated 300 MW. For comparison, this is the equivalent of 250,000 U.S. households \citep{eia2024electricity}.\footnote{10,800 kWh /8760 h = 1.23 kW; 312 MW/ 1.23 kW = 250,000}

The rapid increase in power required for training frontier models is well documented \citep{fist2024energy,sevilla2024aiscaling,pilz2025power}. We discuss whether this trend can continue in Section \ref{sec:Discussion:Can_trends_continue:Power_requirements_2030}.
\vspace*{3mm}
{\setlength\intextsep{0pt}
\begin{figure}[H]
    \centering
    \includegraphics[width=1\linewidth]{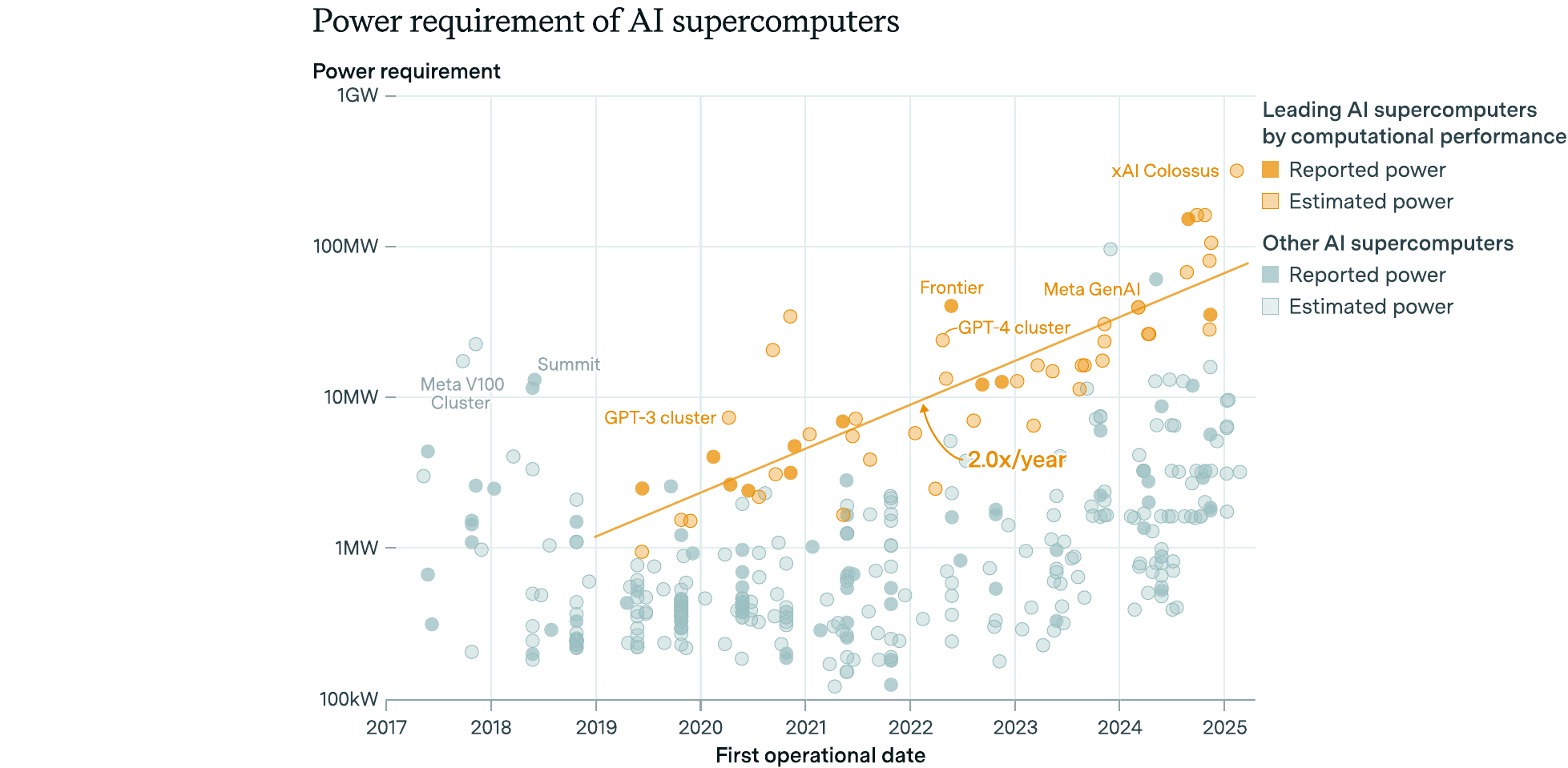}
    \caption{Peak data center power requirements of leading 10 AI supercomputers doubled every year (90\% CI: 1.6--2.2$\times$ per year). We display reported power requirements whenever available. If not, we estimate capacity based on the number and type of chips used.}
    \label{fig:Power_capacity_scatterplot_since_2017}
\end{figure}
}
\subsubsection{Energy efficiency of the leading AI supercomputers improved by 1.34$\times$ per year}
\label{sec:Results:Power_capacity:Energy_efficiency}

We calculate AI supercomputer energy efficiency in FLOP/s per watt (16-bit precision), including both hardware and data center power needs. To calculate efficiency, we divide the computational performance in FLOP/s by the reported or estimated data center power requirement in watts. Energy efficiency at the data center level includes servers, additional cluster components like networking switches, and supporting infrastructure like cooling and power conversion.

We find that between 2019 and 2025, AI supercomputer energy efficiency improved by 1.34$\times$ every year (Figure \ref{fig:energy_efficiency_scatterplot_since_2019}). Holding computational performance constant, AI supercomputers required about 25\% less energy per year.\footnote{We are measuring energy efficiency as peak theoretical FLOP/s divided by required peak power for the AI supercomputer. This is different from energy efficiency in practice, which will be realized FLOP/s divided by average power consumption.} This is roughly in line with the 1.31$\times$ annual increase in energy efficiency of the most efficient supercomputers in the Top500 across the study period in \citet{benhari2024survey}.\footnote{\citet{benhari2024survey} report a maximum value of $4.5\times10^{9}$ FLOP/s per watt for 2013 and $6.5\times10^{10}$ FLOP/s per watt in 2023, implying a 1.31$\times$ annual increase. Note that \citet{benhari2024survey} report the energy efficiency of the most energy‑efficient systems, whereas we report the energy efficiency of the most performant systems. However, the median in Figure 5 of their paper seems to track the maximum efficiency closely, implying that this trend is likely consistent throughout their data, including for the top‑10 most performant systems.}

\vspace{2mm}
{\setlength\intextsep{0pt}
\begin{figure}[H]
    \centering
    \includegraphics[width=1\linewidth]{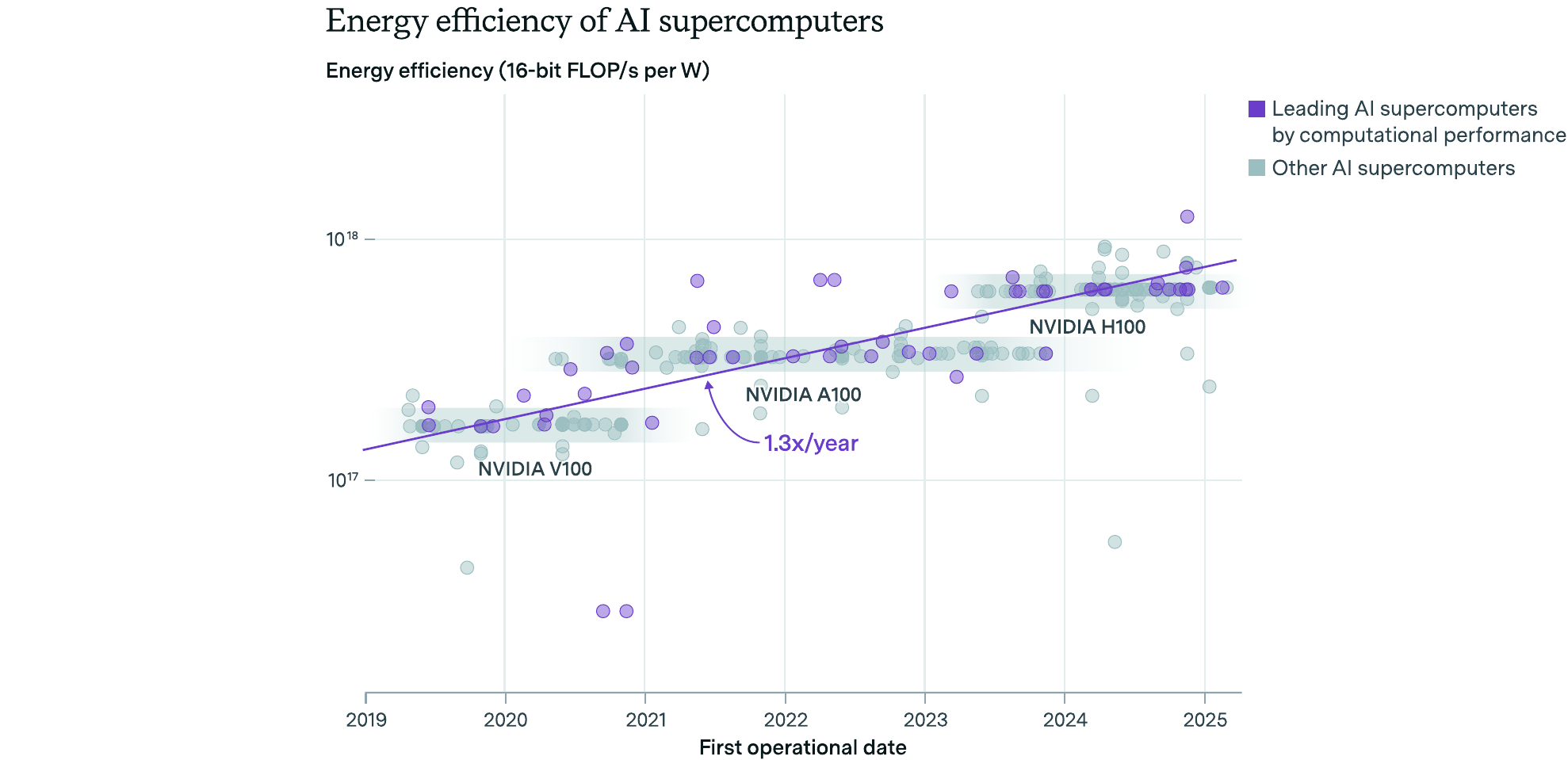}
    \caption{The energy efficiency (16-bit FLOP/s per watt) of top-10 leading AI supercomputers increased by 1.34$\times$ per year between 2019 and 2025 (90\% CI: 1.25--1.43$\times$). Adoption of new chips was the primary driver of energy efficiency improvements, with data center infrastructure efficiency only playing a minor role. We use reported power requirement whenever available and estimated power otherwise. For detailed methods and limitations, see Appendix \ref{sec:Appendix:Methods:Power_Requirements}. 
}
    \label{fig:energy_efficiency_scatterplot_since_2019}
\end{figure}
}

Energy efficiency improvements for AI supercomputers can come from two sources: improvements in hardware efficiency and efficiency improvements in the data center infrastructure, such as cooling. Hardware efficiency improvements primarily stem from improvements in the AI chips, but also include improvements in other hardware such as CPUs, network switches, and storage.\footnote{Note our estimation assumes a fixed ratio of AI chip power requirement to total IT power requirement and thus does not account for efficiency improvements in AI supercomputer components that are independent of efficiency improvements in AI chips.} We model improvements in the energy efficiency of the data center hosting the AI supercomputer by assuming they follow industry-wide trends in Power Usage Effectiveness (PUE) reported by \citet{shehabi2024data}. PUE is the quotient of power supplied to hardware divided by power supplied to the data center. An ideal PUE of 1.0 would indicate that all power delivered to the data center goes directly to the hardware and no power is lost in voltage conversion or needed for cooling and other operations \citep{pilz2023estimating}.

Figure \ref{fig:energy_efficiency_scatterplot_since_2019} shows significant improvements in energy efficiency each time a new AI chip becomes available. Meanwhile, PUE has improved more slowly and was already close to the ideal value of 1.0 in our estimate, causing efficiency improvements of less than 5\% each year \citep{shehabi2024data}. Thus, energy efficiency improvements primarily resulted from AI supercomputers adopting more energy-efficient hardware.

\subsection{The hardware cost of the leading AI supercomputers doubled every year} 
\label{sec:Results:Cost}

We analyze annual growth in the hardware cost for leading AI supercomputers based on either publicly reported cost figures or---if those are unavailable---by estimating the total hardware cost, based on the quantity of chips used and publicly available price data. We further include the estimated cost of additional hardware such as CPUs and network switches, but we do not model power generation or data center construction costs. We apply an inflation adjustment to all values to show costs in January 2025 dollars. Our cost estimates significantly diverge from the values reported by owners, but this could be because reported values primarily come from public projects that often get higher discounts on hardware purchases.\footnote{We estimate our hardware cost data is within 3$\times$ the actual hardware cost in 90\% of cases. See Appendix \ref{sec:Appendix:Methods:Power_Requirements:Limitations} for a longer discussion of limitations and precision of our cost estimates.}

We find that the hardware cost of the leading AI supercomputers increased by 1.9$\times$ every year between 2019 and 2025. Our limited pre-2019 data indicates that hardware costs of more than \$100 million were not uncommon before our study period, with Oak Ridge National Lab's Summit costing about \$200 million in 2025 USD. The most expensive AI supercomputer as of March 2025 was xAI’s Colossus with an estimated hardware cost of \$7 billion.
\vspace{3mm}
{\setlength\intextsep{0pt}
\begin{figure}[H]
    \centering
    \includegraphics[width=1\linewidth]{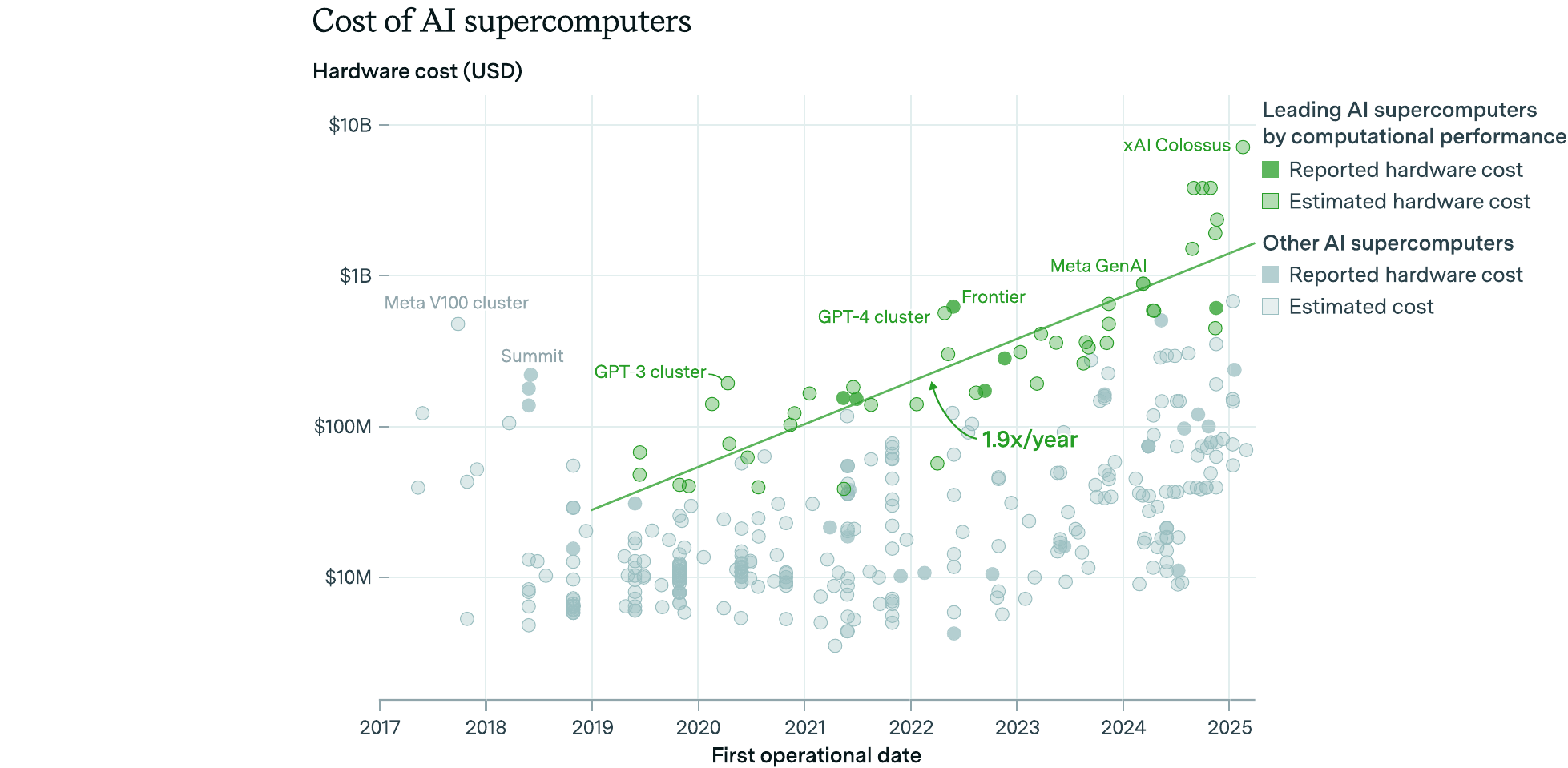}
    \caption{The hardware cost of leading AI supercomputers (by 16-bit performance) when first operational has grown at a rate of 1.9$\times$ (90\% CI: 1.8--2.1$\times$ per year) from 2019 to 2025. We use reported costs or, if unavailable, model costs using a hardware acquisition cost formula adapted from \citet{cottier2024historical}. We adjust all values for inflation to show 2025 USD. For a full explanation of the methods, see Appendix \protect\ref{sec:Appendix:Methods:Cost}}
\end{figure} 
}
The 1.9$\times$ annual growth in hardware costs for leading AI supercomputers is slower than the 2.4$\times$ (90\% CI: 2.0--2.9$\times$) annual increase in total training costs reported by \citet{cottier2024historical}. This difference is due to two factors: First, training durations for frontier models have been extending by 1.4$\times$ annually \citep{frymire2024evolution}, meaning training runs use the same AI supercomputer for longer, which increases the amortized cost even if the AI supercomputer cost stays the same. Second, research personnel costs are a substantial and increasing fraction of AI development, but do not impact the hardware cost of AI supercomputers \citep{cottier2024historical}.

\subsection{Limitations of our data coverage}
\label{sec:Results:Limitations}

Before analyzing the distribution of AI supercomputers across sectors and countries, we emphasize two important limitations in our dataset:
\begin{itemize}[topsep=0pt]
    \item[a)] \textbf{We only capture between 10 and 20\% of all AI supercomputers that fall within our definition.} Specifically, we estimate our dataset covers about 10\% of all relevant AI chips produced in 2023 and 2024 and about 15\% of the chip stocks of the largest companies at the start of 2025. Our dataset covers about half of all systems used in the 25 largest training runs in \citet{epochai2025notable} as of March 2025. The low coverage means our data has limited precision, and a single system being added can significantly change the overall distribution.
    \item[b)] \textbf{The level of coverage likely significantly varies across sectors, chip types, and companies.} For instance, we capture about half of Meta’s total AI supercomputer performance while we capture none of Apple’s AI supercomputers. We also likely cover government AI supercomputers much better than industry systems, since governments tend to be much more transparent about their projects.
\end{itemize}

Given these limitations, we focus on the distribution of AI supercomputers across sectors and countries because both provide reliable insights despite our low coverage: The shift in ownership from public to private sector is a large and robust effect across our entire dataset. Our country-level data is likely robust because we were able to cross-check it against other data (see Appendix \ref{sec:Appendix:Limitations:Comparing_to_public_reports}). Meanwhile, we do not analyze distributions across specific AI chip types or individual companies, as these would be more susceptible to the coverage biases in our dataset.

\subsection{Companies now own the majority of AI supercomputers}
\label{sec:Results:Distribution_by_private_vs_public_sector}

For each AI supercomputer in our dataset, we classify the owner into one of three categories:
\begin{itemize}[topsep=0pt]
    \item[{\textbullet}]Private: The owner is a company.
    \item[{\textbullet}]Public: The owner is a government entity or a university.
    \item[{\textbullet}]Public/Private: The AI supercomputer has several owners belonging to both sectors or if a private project received more than 25\% of the total funding from a government.
\end{itemize}

We find that the share of private sector compute rapidly increased from less than 40\% in 2019 to about 80\% in 2025 (Figure \ref{fig:distribution_of_supercomputers_public_vs_private_sector}), while the share of public AI supercomputers rapidly decreased from about 60\% in 2019 to about 15\% in 2025.\footnote{The figure below shows the trend in 16-bit precision. When considering performance across precisions, the trend is similar, with private owners making up about 85\% of all AI supercomputers in 2025.} Our data may even underestimate this shift, given that companies are less likely to publish data on their systems than public owners. However, note that public sector entities may still be able to access private sector AI supercomputers, given many are available through cloud services. In Section \ref{sec:Discussion:Supercomputer_growth_relied_on_and_enabled_AI_development} we discuss how increased economic importance of AI development and deployment likely led to the rapid increase in private sector share.
\vspace{1mm}
{\setlength\intextsep{0pt}
\begin{figure}[H]
    \centering
    \includegraphics[width=1\linewidth]{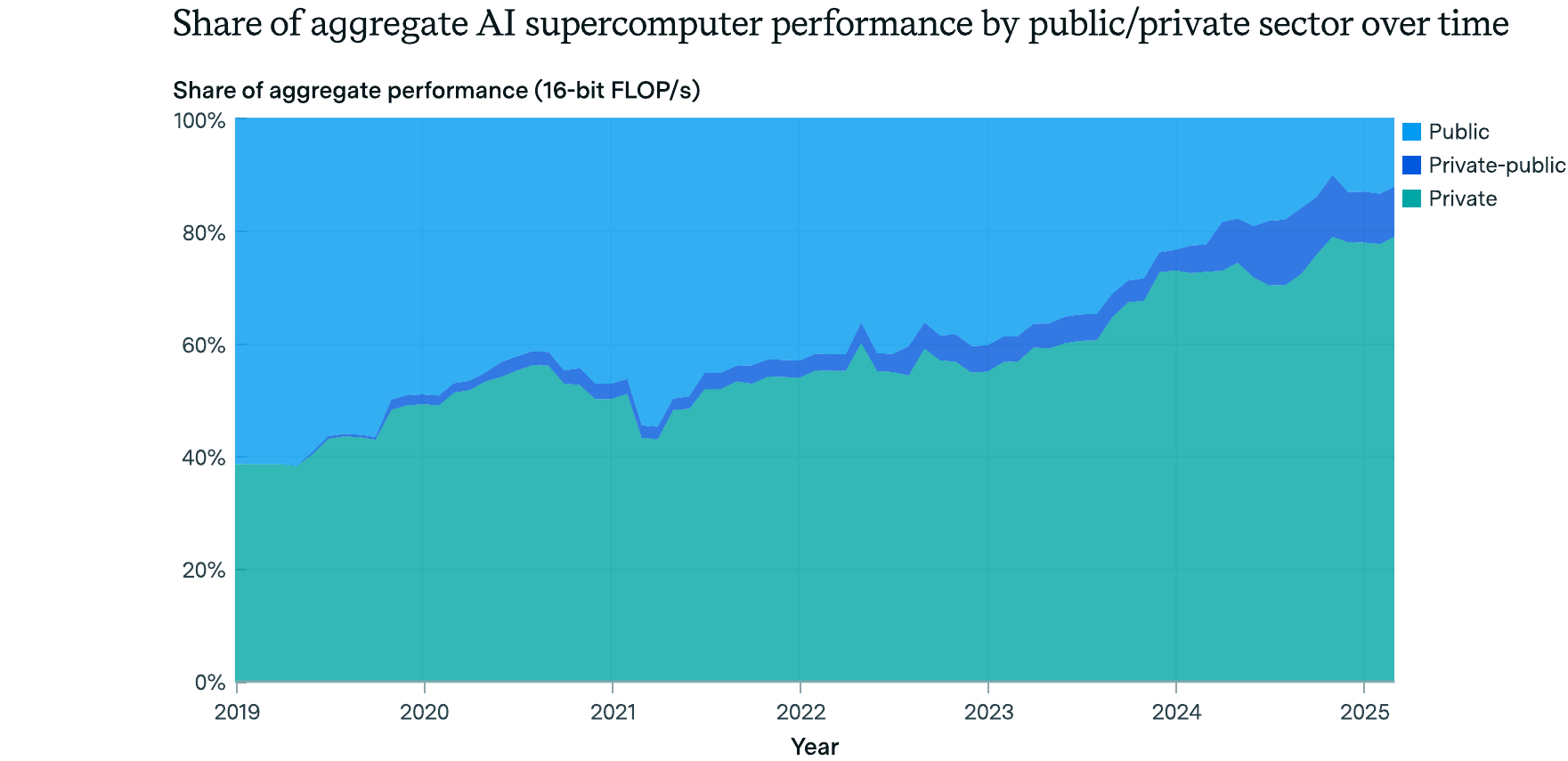}
    \caption{Relative performance shares of public and private sectors based on the owner of the AI supercomputer. An AI supercomputer may have several owners (e.g., if it was a collaborative project or if a government funded an industry project).}
    \label{fig:distribution_of_supercomputers_public_vs_private_sector}
    \end{figure}
}

\subsection{The United States accounts for the majority of global AI supercomputer performance, followed by China}
\label{sec:Results:Distribution_by_country}

When analyzing the distribution across countries, we find that the United States accounted for about 70\% of the computational performance in our dataset at the start of 2019, while China accounted for about 20\% (Figure \ref{fig:distribution_of_supercomputers_by_country}).\footnote{Note that physical location of an AI supercomputer does not directly determine access, given many of our systems are available through cloud services. Furthermore, location also does not necessarily determine ownership since AI supercomputers sometimes belong to owners headquartered in other countries.} Between 2019 and 2022, the Chinese share grew considerably, reaching about 40\% at the start of 2022, although we are unsure if this reflects a real trend or is an artifact of our low data coverage. China's share has since diminished; in March 2025, the United States hosts around 75\% of AI supercomputers by performance while China has around 15\%.

{\setlength\intextsep{0pt}
\begin{figure}[H]
    \centering
    \includegraphics[width=1\linewidth]{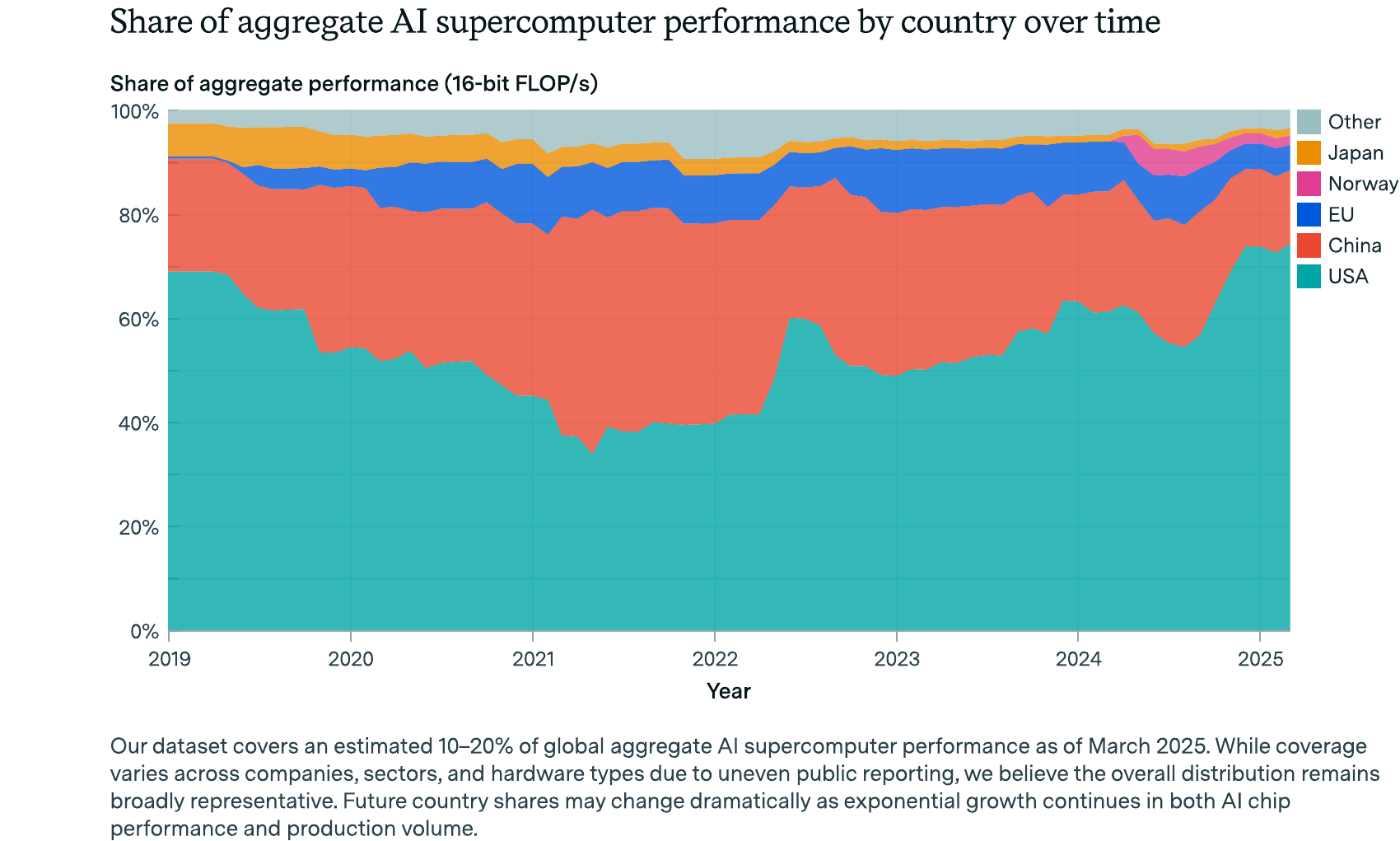}
    \caption{Share of aggregate 16-bit computing power by country over time from AI supercomputers in our dataset. We are visualizing all countries that held a more than 3\% share at some point in time. See Appendix \ref{sec:Appendix:Limitations:Summary:Coverage_percentage} for a discussion of our data coverage.}
    \label{fig:distribution_of_supercomputers_by_country}
\end{figure}
}
As of March 2025, all operational U.S.-based AI supercomputers in our dataset have a combined performance of 850,000 H100-equivalents ($9.1\times10^{20}$ FLOP/s), followed by China with 110,000 H100-equivalents ($1.9\times10^{20}$ FLOP/s) and the European Union with 50,000 H100-equivalents ($5.6\times10^{19}$ FLOP/s) (Figure \ref{fig:aggregate_compute_by_country_bar_chart}). Total computational performance in the United States is thus almost 9 times larger than in China and 17 times larger than the total performance in the European Union.
\vspace{2mm}
{\setlength\intextsep{0pt}
\begin{figure}[H]
    \centering
    \includegraphics[width=0.77\linewidth]{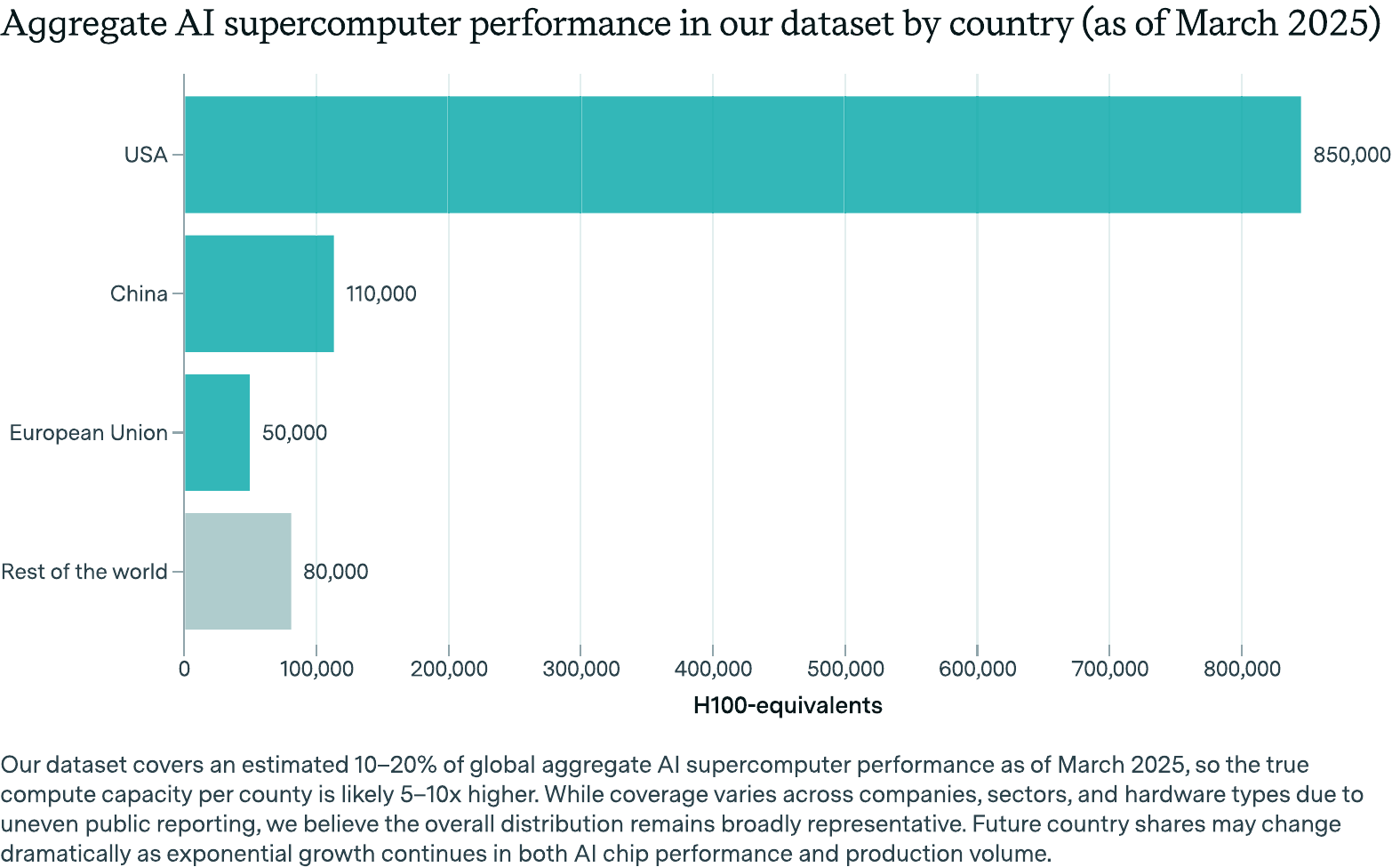}
    \caption{Total AI supercomputer performance by country in H100-equivalents. To convert a system's performance to H100-equivalents, we first take the performance in the lowest precision its AI chips support, considering 32-bit, 16-bit, and 8-bit. We then divide by the 8-bit performance of the H100.
}
    \label{fig:aggregate_compute_by_country_bar_chart}
\end{figure}
}

\section{Discussion}
\label{sec:Discussion}

In this section, we first discuss what caused the rapid growth of AI supercomputer performance and resource needs. We then extrapolate these trends until 2030 and briefly discuss whether the growth in number of chips, power, and hardware cost can continue. We further discuss the geopolitical implications of AI supercomputer distribution across countries and how the increased industry share of AI supercomputers may impact AI research.

\subsection{Rapid growth in AI compute both relied on and enabled the increasing economic importance of the AI industry}
\label{sec:Discussion:Supercomputer_growth_relied_on_and_enabled_AI_development}

The rapid growth in AI supercomputer performance we observe has been primarily driven by a surge in AI investment. While traditional improvements in chip design and manufacturing have contributed to this growth \citep{roser2023moores, hobbhahn2023progress}, AI supercomputers have grown much faster than traditional AI supercomputers (Section \ref{sec:Results:Computational_performance:AI_supercomputers_faster_than_traditional_supercomputers}). This acceleration reflects a fundamental shift in the primary use case of AI supercomputers from academic tools for scientific discovery to industrial machines running economically valuable workloads.

In 2019, the largest AI supercomputers were dominated by government supercomputers like the U.S. Department of Energy's Summit and Sierra. These systems were designed to handle a variety of workloads across different scientific domains and advance foundational research \citep{ornl_summit}. However, in the early 2020s, companies increasingly used AI supercomputers to train AI models with commercial applications, such as OpenAI's GPT-3 and GitHub's Copilot integration \citep{brown2020language, github_copilot_2021}. These demonstrations of AI capabilities led to a significant increase in investment in AI, creating a record demand for AI chips \citep{ourworldindata2024,samborska2024nvidia,richter2025semiconductor}.

As investments in AI increased, companies were able to build more performant AI supercomputers with more and better AI chips. This created a reinforcing cycle: increased investment enabled better AI infrastructure, which produced more capable AI systems, which attracted more users and further investment. The growth of AI supercomputers was therefore both a result of increased funding and a cause of continued investment as AI supercomputers demonstrated their economic value.

\subsection{Can the observed trends continue?}
\label{sec:Discussion:Can_trends_continue}

In Section \ref{sec:Results:Computational_performance:Supercomputer_performance_vs_training_run_required_performance}, we conclude that AI supercomputers have kept pace with the 4--5$\times$ annual growth in compute in the largest AI training runs.\footnote{At least when considering only industry AI supercomputers and considering performance across precisions.} This section will discuss what it would mean for each of the trends in chips, hardware cost, and power needs to continue until 2030.\footnote{Our extrapolations do not model deviations from the current training duration growth. If the duration of the largest AI training runs continues to increase by 1.4$\times$ annually, the largest training runs in 2030 may last 16 months, exceeding the optimal training duration according to \citep{epochlongest}. If training run duration stops increasing, AI supercomputers would have to grow at a faster rate to sustain a 4--5$\times$ annual increase in training compute for the largest AI models.}
{\setlength\intextsep{0pt}
\begin{table}[]
\caption{Historical data and extrapolation of trends based on the current largest AI supercomputer and historical growth rates described in Section \ref{sec:Results}. Using the growth rates of only industry-owned AI supercomputers would lead to higher extrapolated values. Extrapolated values are rounded to not imply precision.}
\label{tableforecasts}
\begin{flushleft}
\begin{small}
\begin{sc}
\begin{tabular}{llrrrrr}
\hline
\abovespace
\textbf{Date} & \textbf{Leading AI } & \textbf{Performance } & \textbf{H100-eq} $^\dagger$& \textbf{Number of }& \textbf{Power} & \textbf{Hardware Cost}  \\
\belowspace
& \textbf{Supercomputer} & (16-bit FLOP/s) & & \textbf{AI chips}& & (2025 USD)\\
\hline
\abovespace
June 2019 & Oak Ridge Summit & $3.46\times10^{18}$ & 3,492 & 28k& 13MW & \$200M \\
June 2020 & Oak Ridge Summit & $3.46\times10^{18}$ & 3,492 & 28k& 13MW & \$200M \\
June 2021 & Sunway OceanLight & $5.95\times10^{18}$ & 6,008 & 108k& N/A & N/A \\
June 2022 & Oak Ridge Frontier & $1.44\times10^{19}$ & 14,566 & 38k& 40MW & \$600M \\
June 2023 & Oak Ridge Frontier & $1.44\times10^{19}$ & 14,566 & 38k& 40MW & \$600M \\
June 2024 & Meta GenAI 2024a & $2.43\times10^{19}$ & 24,576 & 25k& 40MW& \$900M \\
March 2025 & xAI Colossus & $1.98\times10^{20}$ & 200k & 200k& 300MW& \$7B \\
June 2026 & \textit{Extrapolated} & $5\times10^{20}$ & 500k & 300k& 600MW & \$14B \\
June 2027 & \textit{Extrapolated} & $1\times10^{21}$ & 1M & 500k& 1GW & \$25B \\
June 2028 & \textit{Extrapolated} & $3\times10^{21}$ & 3M & 800k& 2GW & \$50B \\
June 2029 & \textit{Extrapolated} & $8\times10^{21}$ & 8M & 1.3M& 5GW & \$100B \\
\belowspace
June 2030 & \textit{Extrapolated} & $2\times10^{22}$ & 20M & 2M& 9GW & \$200B \\
\hline
\end{tabular}
\end{sc}
\end{small}
\scriptsize
\begin{list}{}{
  \setlength{\leftmargin}{0.2in}
  \setlength{\rightmargin}{0.2in}
}
\item[$^\dagger$] Here, we define H100-equivalents (H100-eq) as the AI supercomputer's 16-bit performance divided by the H100's 16-bit performance. This is different than elsewhere in the paper, where we defined it in terms of maximum performance over 8, 16, or 32 bit. H100-equivalents is not a standardized measurement, and should be used only to get a general sense of the scale.
\end{list}
\end{flushleft}
\vskip -0.1in
\label{table:extrapolation_of_trends_by_year}
\end{table}
}
\subsubsection{The largest AI supercomputer could need two million chips by 2030}
\label{sec:Discussion:Can_trends_continue:Chip_requirements_2030}

If the number of AI chips continues increasing by 1.6$\times$ every year, the largest AI supercomputer in 2030 will require about 2 million AI chips (Table \ref{tableforecasts}). \citet{sevilla2024aiscaling} estimated that AI chip production could increase by 1.3$\times$ to 2$\times$ annually until 2030. Extrapolating from present-day chip production\footnote{Public sources estimate that NVIDIA shipped about 500k H100s in 2023 and 2 million in 2024, for a total of 2.5 million H100s \citep{nolan2023nvidia,shilov2023nvidia}. However, some analysts \citep{tweaktown} estimate NVIDIA produced up to 1.5 million H100s in Q4 of 2024. Assuming NVIDIA produced about 1M H100s on average per quarter in 2024 yields a total of 4.5 million H100s. NVIDIA produces the majority of all AI chips \citep{sastry2024compute}.} this indicates a production of 7.4M to 144M AI chips annually in 2030.\footnote{For the low-end range, we consider an annual production of 2 million AI chips, growing by 1.3$\times$ every year: 2M *1.3$^5$ = 7.4M. For the high-end range, we consider an annual production of 4.5 million AI chips, growing by 2$\times$ every year: 4.5M * 2$^5$ = 144M} If the largest AI supercomputer used 2 million AI chips in 2030, it would need between 1\% and 27\% of global annual AI chip production, indicating this scale is feasible if AI chip production continues growing at the estimated rate.

\subsubsection{The largest AI supercomputer could have a hardware cost of about \$200B by 2030}
\label{sec:Discussion:Can_trends_continue:cost_requirements_2030}

If the hardware cost of the leading AI supercomputers continues to increase at a rate of 1.9$\times$ annually, the leading system’s hardware cost in 2030 will be about \$200B (in 2025 USD). This is in addition to the cost of the data center facility, which is likely about \$10B per GW, adding a further \$90B to the acquisition cost \citep{pilz2023estimating}.

Current AI infrastructure is already close to this scale: In 2025, Microsoft announced plans to spend \$80B on AI infrastructure globally, and AWS announced plans to spend more than \$100B \citep{smith2025infrastructure, gonsalves2025aws}. Meanwhile, OpenAI announced plans to spend up to \$500B on the Stargate project over four years \citep{openai2025stargate}. These announcements are compatible with \$200 billion hardware costs for a single project by 2030, especially as AI investment is projected to continue growing \citep{precedence2025ai,idc2025ai, gvresearch2024}.

\subsubsection{The largest AI supercomputer could need 9 GW of power by 2030}
\label{sec:Discussion:Can_trends_continue:Power_requirements_2030}

If AI supercomputer power requirements continue growing at a rate of 2.0$\times$ every year, the leading AI supercomputer will need about 9 GW of power in 2030 (Table \ref{tableforecasts}). This is slightly higher than \citet{sevilla2024aiscaling}’s extrapolation of 6 GW and matches \citet{pilz2025power}’s estimate for the AI supercomputer running the largest training run in 2030. 

The largest data center campuses today have a capacity of hundreds of MW, and as of early 2025, no existing campus exceeding 1 GW has been publicly reported \citep{pilz2023estimating}. While a 2 GW AI supercomputer in 2028 is likely feasible, a system with a capacity of 9 GW by 2030 would require as much power as 9 nuclear reactors can generate, and would likely face severe permitting and equipment supply chain challenges, as well as other potential challenges such as local community opposition \citep{pilz2025power}.\footnote{In 2025, the U.S. government began programs to support large-scale data center campuses and several companies have already announced plans to build multi-GW data centers \citep{moss2024, swsk2024}. Still, no known industrial facilities currently require several GW of power, indicating that this amount of power may be challenging to secure \citep{sevilla2024aiscaling}.} As they struggle to secure adequate power, companies may increasingly use decentralized training techniques that allow them to distribute a training run across AI supercomputers in several locations. Some notable training runs, including Google DeepMind’s Gemini 1.0 and OpenAI’s GPT-4.5, were reportedly already trained across several AI supercomputers \citep{moss2023google,openai2025stargate,whitehouse2025ai}.

\subsubsection{Conclusion: Power constraints will likely be the main constraint to continued growth} \label{sec:Discussion:Can_trends_continue:conclusion} 

Power constraints will likely become the primary bottleneck for AI supercomputer growth, driving a shift toward distributed training across multiple sites. This evolution could change how we measure AI training capabilities---from focusing on individual AI supercomputers to assessing companies' aggregate compute capacity. While chip production and hardware cost trends appear sustainable through 2030, the continuation of all these trends ultimately depends on AI applications delivering sufficient economic value to justify the massive investments required for infrastructure expansion.

\subsection{U.S. dominance in global AI supercomputer distribution}
\label{sec:Discussion:Geopolitical_competition}

This section discusses that U.S. dominance likely resulted from leading in related industries, and will likely continue, given stated U.S. policy and U.S. control of key AI chip production chokepoints.

\subsubsection{U.S. dominance resulted from dominance in cloud computing and AI development}
\label{sec:Discussion:Geopolitical_competition:US_tech_infrastructure_enables_lead}

According to our data, around 75\% of all AI supercomputer performance is currently based in the United States (Figure \ref{fig:distribution_of_supercomputers_by_country}). How did the United States develop such a dominant position in AI supercomputers, while countries that used to play a prominent role in public supercomputing, like the UK, Germany, or Japan, declined in importance?

U.S. dominance was likely a direct result of AI supercomputers becoming increasingly commercialized and dominated by companies (instead of governments or academia), which were primarily based in the United States due to dominance in previous technologies. This advantage is evident in cloud computing infrastructure, where in 2019, the top three leading U.S. cloud companies, AWS, Microsoft, and Google alone made up 68\% of global market share \citep{gartner2020iaas}. American companies also played leading roles in key AI advances, including in recommender systems, scientific applications like AlphaFold, and LLM chatbots like ChatGPT \citep{dong2022recommender,jumper2024alphafold,openai2022chatgpt}. Overall, American companies were involved in developing 338 of the 476 notable AI models and trained 18 of the 25 largest AI models by training compute recorded by \citep{epochai2025notable}. While limited reliable data on global market shares in AI applications exists, record user growth may indicate that U.S. companies also lead in total number of users \citep{hu2023chatgpt}.

\subsubsection{The United States will likely continue leading in AI supercomputers}
\label{sec:Discussion:Geopolitical_competition:US_will_likely_continue_lead}

The United States dominates not only AI development and cloud provision, but also the design of AI chips, and several inputs to semiconductor manufacturing \citep{sastry2024compute}. The U.S. government has previously used its dominance in AI chips to impose export controls on AI chips and key equipment to China, and introduced an AI diffusion framework that puts conditions on the export of AI chips to countries that are not close U.S. allies \citep{allen2022export, heim2025diffusion}. 

At the same time, some challenges could limit U.S. dominance in AI supercomputers:
\begin{itemize}[topsep=0pt]
    \item[{\textbullet}]\textbf{Power requirements:} AI's power demand is massively increasing, both in terms of the power needed for AI supercomputers and in terms of the overall number of AI chips deployed, primarily for inference. \citep{pilz2023estimating}. The United States is facing significant challenges to add enough power generation capacity to sustain the current rate of AI data center growth \citep{pilz2025power,fist2024energy,mahmood2025grid}.
    \item[{\textbullet}]\textbf{Investment in sovereign infrastructure by foreign governments:} Some governments have begun investing in local AI infrastructure, such as France \citep{reuters2025france}, the United Kingdom \citep{ukdsit2025ai}, Saudi Arabia \citep{benito2024saudi}, and the UAE \citep{allen2025uae}. However, most of these projects are small compared to leading U.S. AI supercomputers. Furthermore, given U.S. control of AI chip production, the United States could block chip access if these projects were threatening U.S. computing dominance.
    \item[{\textbullet}]\textbf{Competition from China:} The Chinese government and Chinese companies are heavily investing in AI infrastructure, but being unable to import leading U.S. AI chips, the country relies on inferior U.S. or domestically produced AI chips. Limited AI chip access makes it more costly to establish significant AI supercomputers, and limits the total number of projects in China \citep{scanlon2025chinese,lin2025chinese}. So far, Chinese efforts at indigenous production of AI chips have been severely hampered by the inability to produce or import crucial equipment like DUV and EUV lithography machines that are extremely challenging to produce \citep{grunewald2023lithography, allen2022export}.
\end{itemize}

To summarize, the United States leads in AI model development and cloud computing and controls key chokepoints in the semiconductor supply chain. Combined with stated government policy to advance U.S. AI leadership, this leads us to conclude that the United States will likely continue leading in AI supercomputers for at least another six years \cite{whitehouse2025ai}.

\subsection{Consequences of increased private sector dominance}
\label{sec:Discussion:Consequences_of_private_sector_dominance}
Our finding that companies own a growing share of AI supercomputers matches a previously reported trend: AI research is increasingly dominated by large companies rather than academic or government groups. \citet{besiroglu2024chinllm} found a stark decline in academic institutions' share of large-scale machine learning models, from approximately 65\% in 2012 to just 10\% in 2023. 

The shift away from public ownership of AI supercomputers is likely due to their increased economic importance (Section \ref{sec:Discussion:Supercomputer_growth_relied_on_and_enabled_AI_development}), which has rapidly increased private AI investments. More investment allowed companies to build systems as expensive as xAI’s Colossus, which had an estimated hardware cost of \$7B. Meanwhile, the most expensive government projects, Frontier and El Capitan, cost only \$600M each. Additionally, governments usually build only a small number of systems for research purposes. Meanwhile, major tech companies often build dozens of AI supercomputers, given that they are not just training larger models, but also serving millions of users around the world.

This shift from public to private ownership of AI supercomputers produces two significant consequences for AI research: restricted access for academic researchers and diminished visibility into AI development and deployment.

\textbf{Limited access for academic researchers: }The concentration of AI supercomputers in industry reduces access to frontier compute resources for academic researchers, who historically have contributed to AI progress and provided independent evaluation and scrutiny \citep{besiroglu2024chinllm}. The ownership of systems does not inherently determine compute access because researchers can rent AI supercomputers through cloud companies \citep{heim2023hardware}. However, renting large quantities of AI chips---beyond a few thousand---for even short durations can still be prohibitively expensive for academic researchers, compelling them to rely on smaller, less powerful models \citep{lohn2023ai}.

\textbf{Lack of visibility:} As companies now operate the leading AI supercomputers, they have become the main driver of frontier AI progress, relegating government and academic labs to a supporting role. Because companies are often less public about their research, governments may increasingly struggle to track capability gains in AI models \citep{besiroglu2024chinllm}. Additionally, given the importance of compute for AI development and deployment, the scale and number of a nation’s top AI supercomputers are increasingly tied to competitiveness in AI. With companies controlling most systems, governments increasingly lack data on the extent of their national AI infrastructure, hampering policymakers’ ability to craft a coherent strategy for technological competition.

One option for governments to increase visibility into AI development and deployment and better understand national competitiveness could be to require companies to report key data about their infrastructure, such as the performance of their largest AI supercomputers and the total extent of their infrastructure \citep{sastry2024compute}. Governments could also collect intelligence on the AI compute capacity of other countries, allowing them to better understand their competitive position and potentially making it easier to verify potential future international agreements on AI \citep{baker2023computing, sastry2024compute}.


\section{Conclusion}
\label{sec:Conclusion}

We compiled a dataset of 500 AI supercomputers between 2019 and 2025 and found that performance, number of chips, power requirements, and hardware cost have all grown exponentially. The rapid performance growth of AI supercomputers, combined with increasing training durations, has enabled a 4--5$\times$ annual increase in training compute for frontier AI models, which has fueled significant advances in AI capabilities and driven further investment in infrastructure. If trends continue, the leading AI supercomputer in 2030 could have a hardware cost of more than \$200 billion and incorporate over 2 million AI chips. However, the projected power requirement of 9 GW would be challenging to secure in a single location, likely forcing companies to adopt decentralized training approaches across multiple sites.

Our data also reveals key trends in AI supercomputer ownership, with companies increasing their share of total AI supercomputer performance from 40\% in 2019 to more than 80\% in 2025. This finding emphasizes the previously observed increasing compute divide between industry and academia. The United States hosts approximately 75\% of global AI supercomputer performance and will likely maintain this dominance through its control over the AI chip supply chain.

To conclude, AI supercomputers have been a key driver of AI progress and represent a central component of the AI supply chain \citep{sastry2024compute}. Our analysis provides valuable information about AI supercomputers' growth patterns, distribution, and resource requirements. Such information will be increasingly important for policymakers, and more generally for understanding the trajectory of AI.


\section*{Acknowledgements}
\label{sec:Acknowledgements}

We would like to thank the following people for their assistance, feedback, and contributions:
\begin{itemize}[topsep=0pt]
    \item David Owen for reliable guidance on scope and execution of this project as well as repeated feedback on the report.
    \item Qiong Fang and Veronika Blablová for substantial contributions to data collection.
    \item Lovis Heindrich, Terry Wei, David Atanasov for assistance with data entry and verification.
    \item Robert Sandler for figure design and Edu Roldán for figure editing. 
    \item Luke Frymire for his work estimating power requirements for AI supercomputers and Ben Cottier for his work estimating hardware acquisition costs of AI supercomputers.
    \item Pablo Villalobos for reviewing our code.
    \item Caroline Falkman Olsson and Jessica P. Wang for typesetting
    \item Various people who reviewed our data and suggested additional systems to include.
    \item The Epoch AI team and everyone else who provided feedback and helpful discussions.
\end{itemize}

\clearpage

\newpage
\abovespace
\addcontentsline{toc}{section}{APPENDIX}
{\centering\Large\bfseries APPENDIX\par}
\appendix

\section{Review of existing data sources}
\label{sec:Appendix:Review_of_existing_data}

\subsection{The Top500 list and its limitations for AI supercomputers}
\label{sec:Appendix:Review_of_existing_data:Top500}

The \href{https://www.top500.org/project/}{Top500 list} has been the primary leaderboard for tracking supercomputer performance since its inception in 1993. It ranks systems based on their performance in solving linear equations using the LINPACK benchmark \citep{dongarra1987linpack}. While this benchmark has provided a consistent, long-term method for comparing traditional high-performance computing (HPC) systems, it has several significant limitations when applied to AI supercomputers:
\begin{itemize}[topsep=0pt]
    \item Participation in the Top500 list is voluntary, leading to significant gaps in reporting. Companies, particularly cloud providers, which own many of the largest AI supercomputers, face limited incentives to report their AI supercomputers. Running the LINPACK benchmark diverts valuable supercomputer and engineer time from more economically valuable uses like AI training or deployment. Instead of reporting to the Top500, companies sometimes independently publish promotional blog posts about their systems \citep{langston2020, meta2022, AWS2023}, while often maintaining ambiguity about the number and size of their largest systems to avoid giving competitors unnecessary information about their strategies. Additionally, Chinese owner stopped reporting any systems to the Top500 list in 2022, presumably to reduce scrutiny and avoid U.S. sanctions \citep{shah2024}.
    
    \item LINPACK is not an AI benchmark. It measures performance on linear equations requiring high-precision 64-bit number formats \citep{dongarra1987linpack}, while modern AI workloads run on much lower precision formats (16-bit, 8-bit, or even 4-bit for some inference workloads\footnote{Or even 4-bit precision for some inference workloads \cite{ashkboos2023quik}.}). While performance on different precision formats was formerly highly correlated, the introduction of tensor cores for lower precision formats on AI accelerators led to drastically faster performance increases in these formats \citep{hobbhahn2023progress,rahman2024trends}. This divergence means LINPACK performance does not accurately capture a supercomputer's performance for AI workloads.\footnote{For instance, Microsoft's Eagle and Japan's Fugaku have comparable performances on LINPACK ($5.6\times10^{17}$ FLOP/s vs $4.4\times10^{17}$ FLOP/s), but given that Fugaku does not contain any GPUs or other chips optimized for low-precision performance, they diverge by almost an order of magnitude on FP8 performance ($2.9\times10^{19}$ FLOP/s vs $4.3\times10^{18}$ FLOP/s) \citep{servethehome, fugakuspecs}.} New benchmarks like HPL-MxP and ML-Perf better capture AI-relevant performance but have not been widely adopted \citep{hplmxp,mattson2019mlperf}.
\end{itemize}
Besides the Top500, no major datasets of supercomputers exist, meaning that previous analyses of supercomputers, such as \citet{hoefflinger2020trends}, \citet{tekin2021analysis} and \citet{chang2024hpc} have exclusively relied on the Top500 list. While these analyses offer useful insights into changes in components, performance, and energy efficiency of traditional supercomputers, the limitations of the Top500 lists discussed above mean the observed trends do not adequately capture AI supercomputers.

\subsection{Commercial databases of AI supercomputers}
\label{sec:Appendix:Commercial_databases}

Some analysts like SemiAnalysis and The Information have private databases of AI supercomputers that are available for paid subscribers. Furthermore, some companies such as Omdia offer trackers of AI chip shipments \citep{semianalysundateddatacentermodel,theinformation2025datacenter,omdia2025datacenter}.
These databases are typically focused on providing business intelligence. Thus, they do not assess historical trends and may not capture data from non-industry sources. Furthermore, these databases usually do not disclose their methods and sources and do not make the analysis of their data publicly available.

\section{Detailed Methods}
\label{sec:Appendix:Methods}

\subsection{Data collection process}
\label{sec:Appendix:Methods:Data_collection_process}

We relied on systematic Google searches and publicly available datasets to find potential AI supercomputers. For each potential AI supercomputer, we conducted an additional search to find and verify all relevant publicly available data about it.

\textbf{Search methodology:}
    \begin{itemize}[topsep=0pt]
        \item[a)] We used the Google Search API to search for terms such as “AI supercomputer” and “GPU cluster” in consecutive 12-day windows (1-1-2019--1-3-2025). We additionally conducted year-by-year country searches (e.g., “Albania AI supercomputer”).
        \begin{itemize}
            \item[{\textbullet}] Although our study period begins in 2019, we also conducted a similar, pared-down Google search for January 2016--January 2019 in order to be able to determine which AI supercomputers were in the top 10 by computational performance at the start of 2019. For this, we reduced our search terms by roughly 80\% to lower the number of records to look through. 
        \end{itemize}
        \item[b)] We parsed the top results with the Beautiful Soup Python package and used GPT-4o via the OpenAI API to extract system names and chip counts of any AI supercomputers mentioned.
        \item[c)] We grouped entries by name in a spreadsheet, deduplicated, verified all potential AI supercomputers manually, and added those that fit our inclusion criteria to our dataset.
        \item[d)] Find additional details about the Google Search methods in Section \ref{sec:Appendix:Methods:Google_search}.
    \end{itemize}
\textbf{Additional sources:}
    \begin{itemize}[topsep=0pt]
        \item[a)] \href{https://www.top500.org/}{Top500 list}, inferring AI chip counts from reported accelerator cores.
         \begin{itemize}
            \item[{\textbullet}] Many systems in the Top500 did not contain AI chips; however, those that did usually listed the ‘Accelerator/Co-Processor' type and the total number of 'Accelerator/Co-Processor Cores.' Since we knew the number of cores for each AI chip model, we calculated the implied AI chip count for the system by dividing the number of cores by the cores per AI chip. We verified this method by checking it for AI supercomputers in the Top500 with previously known AI chip counts.
            \item[{\textbullet}] We considered all Top500 entries from June 2014 to November 2024 (but included only those that qualified for our inclusion criteria between 2017 and 2025).
         \end{itemize}
        \item[b)] Epoch AI’s \href{https://epochai.org/data/notable-ai-models-documentation}{notable AI models dataset}.
        \item[c)] Published compilations of Chinese AI supercomputers (redacted, please reach out).
        \item[d)] A small number of entries from a project on sovereign compute resources led by Aris Richardson (publication forthcoming).
        \item[e)] \href{https://mlcommons.org/benchmarks/training/}{MLCommons Results}.
        \item[f)] \href{http://gpulist.ai}{gpulist.ai} (last accessed January 2025).
        \item[g)] Articles and newsletters shared by colleagues, such as from \href{https://semianalysis.com/}{SemiAnalysis}, \href{https://www.transformernews.ai/}{Transformer}, and \href{https://importai.substack.com/p/import-ai-375-gpt-2-five-years-later}{Import AI}.
    \end{itemize}
\textbf{Remaining components:}
    \begin{itemize}[topsep=0pt]
        \item[{\textbullet}] We built our initial dataset via Google Alerts for the keyword "AI supercomputer" (June 2023--Aug 2024)\footnote{Rose Hadshar and Angelina Li contributed additional entries during the Epoch FRI Mentorship Program 2023.}.
        \item[{\textbullet}] Two Chinese-language analysts conducted targeted searches of systems in China and Hong Kong (see Appendix \ref{sec:Appendix:Methods:Chinese_supercomputers}).
        \item[{\textbullet}] Our main data collection focused on AI supercomputers that first became operational between 2019 and 2025. However, we also included AI supercomputers that became operational between 2017 and 2019 if they met the standard inclusion criteria, or if they were operational before 2017 and were at least 1\% as large as the largest known supercomputer in January 2017.
        \item[{\textbullet}] We collected various additional sources for details on specific supercomputers using the Perplexity API.
        \item[{\textbullet}] For over 500 key supercomputers, an Epoch staff member did an additional verification of the entry (marked as true in the 'Verified Additional Time' field). This focused on systems that were especially large for their time, most Chinese systems, and any outliers.
    \end{itemize}

A full, up-to-date documentation of all fields in our dataset is available at:
\url{https://epoch.ai/data/ai-supercomputers-documentation}

\subsection{Google search methodology}
\label{sec:Appendix:Methods:Google_search}

We conducted automated Google searches spanning from January 2019 to March 2025 for consecutive 12-day windows, using various keywords related to AI supercomputers. For each search term, we collected different amounts of results based on their utility in finding relevant information:
\begin{itemize}[topsep=0pt]
    \item ``AI Supercomputer'': 30 Google results
    \item ``AI Supercomputer cluster'': 30 Google results
    \item ``AI Supercomputer news'': 20 Google results
    \item ``AI Supercomputer cluster news'': 20 Google results
    \item ``GPU Cluster'': 20 Google results
    \item ``Compute Cluster'': 10 Google results
    \item ``V100 Cluster'': 10 Google results
    \item ``A100 Cluster'': 10 Google results
    \item ``H100 Cluster'': 10 Google results
\end{itemize}
We parsed all websites using the \citet{BeautifulSoup} Python library and used GPT-4o from the OpenAI API to search for information on all mentioned AI supercomputers (see prompt below).

Our searches yielded over 20,000 unique websites, resulting in approximately 2,500 potential AI supercomputer mentions after deduplication. For each unique AI supercomputer, we used the Perplexity API to collect additional data sources (see prompt below). 

\subsubsection*{GPT-4o Prompt for Initial Extraction} \label{sec:chatgptprompt}

\textit{Here is the text from a webpage that potentially contains some information about AI supercomputers. Please list the names of any AI supercomputer clusters that are listed in this article, separated by semicolons if there are multiple. If you know the company/organization name that owns/runs it, you should write the supercomputer name as the company/organization name, followed by the name of the cluster. If the cluster does not have a name, simply refer to it with `UNNAMED' and include any identifiable information given. Please include any information about the number and type of AI chips (e.g. GPUs or TPUs) in square brackets after the cluster name. Say `[NOINFO]' if there is no information in the article about chip type or quantity. For example, a response might look like `OpenAI Stargate [NOINFO]; Frontier [37,632 AMD MI250X]; Microsoft UNNAMED Arizona H100s [50,000 NVIDIA H100s]'. You should only list AI supercomputer clusters and associated chip information, nothing else. If there are no supercomputer clusters mentioned in the article, just reply with `None'. If you can't access or read the article, just reply with `Could not access article'. However, this should be rare, and mainly only happen if the article is paywalled. Do not mention any other details. Article text: \{TEXT HERE\}}

\subsubsection*{Perplexity Prompt for Detailed Information} \label{sec:perplexityprompt}

\textit{Tell me all the details you can about the \{SUPERCOMPUTER NAME\} supercomputer, including but not limited to: What type of AI accelerator chips (eg GPUs, TPUs, etc) do they use (be as specific about the exact type of chip as possible)? How many do they have, if any? When was it completed, or when is it expected to be completed? When was it first announced? What is the timeline for any updates/iterations to this supercomputer? Where is it located? (be as specific as possible) How many AI FLOP/s could it do? Who operates it? Who uses it? Who owns the supercomputer? Please list several organizations if it is a joint partnership, and list if these organizations are or part of government, academia, industry, or something else? Are there multiple supercomputers that could go by roughly this name? Have there been different versions/iterations of this supercomputer?}

\subsection{Approach for finding Chinese AI supercomputers}
\label{sec:Appendix:Methods:Chinese_supercomputers}

We decided to redact our approach to finding Chinese AI supercomputers and avoid providing identifying information about them throughout the paper to preserve data sources. We take this step as a precautionary measure because Chinese websites cited in public reports have been redacted or replaced with malware in the past \citep{wsj2023lid}.

If you would like to request access to our methodology for Chinese AI supercomputers, please contact Konstantin at kfp15@georgetown.edu.

\subsection{Power requirements}
\label{sec:Appendix:Methods:Power_Requirements}

We calculated the peak power demand for each AI supercomputer with the following formula: 
\vspace{3mm}
\setlength{\abovedisplayskip}{0pt}
\setlength{\belowdisplayskip}{0pt}
\begin{align*}
\textit{Chip TDP} \times \textit{number of chips} \times \textit{system overhead} \times \textit{PUE} 
\end{align*}
We collected Thermal Design Power (TDP) for most chips when publicly available, though we did not find the TDP for some Chinese chips and custom silicon such as Google's TPU v5p. We considered both primary and secondary chips when counting the number and types of chips. We used a 2.03$\times$ multiplier for non-GPU hardware to account for system overhead (additional power needed for other server components like CPUs, network switches, and storage), based on NVIDIA DGX H100 server specifications \citep{nvidia_dgx_h100_user_guide_2025}. We also factored in Power Usage Effectiveness (PUE), which is the ratio of total data center power use to IT power use (with a minimum value of 1). According to the 2024 United States Data Center Energy Usage Report \citep{shehabi2024data}, specialized AI datacenter facilities had an average PUE of 1.14 in 2023, which is 0.29 lower than the overall national average of 1.43. We adjusted the trend for all datacenter facilities to estimate the average PUE of AI datacenters by subtracting 0.29 from the overall values reported by \citet{shehabi2024data} (Table \ref{table:pue_by_year}).

{\setlength\intextsep{0pt}
\begin{table}[H]
\vspace*{3mm}
\caption{AI data center power usage effectiveness (PUE) over time, adapted from \citep{shehabi2024data}.}
\label{sample-table}
\begin{center}
\begin{small}
\begin{sc}
\begin{tabular}{lcccccccccr}
\hline
\abovespace\belowspace
\textbf{Year} & 2016 & 2017 & 2018 & 2019 & 2020 & 2021 & 2022 & 2023 & 2024 & 2025 \\
\hline
\abovespace
\textbf{PUE} & 1.31 & 1.29 & 1.26 & 1.22 & 1.20 & 1.18 & 1.17 & 1.14 & 1.12 & 1.10 \\
\belowspace
\end{tabular}
\end{sc}
\end{small}
\end{center}
\vskip -0.1in
\label{table:pue_by_year}
\end{table}
}
The full formula we use is:
\setlength{\abovedisplayskip}{3pt}
\setlength{\belowdisplayskip}{0pt}
\begin{align*}
\textit{Power} = & \left[ (\textit{Primary AI chip TDP} \times \textit{Primary AI chip quantity})\right. \\
& \left. + (\textit{Secondary AI chip TDP} \times \textit{Secondary AI chip quantity}) \right] \\
& \times \textit{Server overhead factor} \times \textit{Datacenter PUE}
\end{align*}

We base some of the reported power values in our dataset on the top 500 list. However, the list reports average power utilization during the benchmark, rather than peak power requirement. To determine peak power, we compare peak and average power for supercomputers where we have both, find that they differ on average by a factor of 1.5, and scale all the Top500 reported power figures by this factor. We then multiply by the PUE in the given year to find peak power demand for the entire system.

 \subsubsection{Limitations with our power data}
\label{sec:Appendix:Methods:Power_Requirements:Limitations}

We rely on owner-reported power estimates for 15\% of the AI supercomputers in our dataset. These reported figures lack standardization---some may represent only critical IT load at theoretical maximum utilization, while others include complete data center infrastructure overhead (accounting for power conversion losses and cooling requirements).

For the remaining 85\% of systems, we estimate the power requirements as detailed in the previous section. A key limitation of our current approach is the application of a uniform 2.03× multiplier for all chip types to account for additional system hardware. Future analyses would benefit from developing chip-specific overhead multipliers that better reflect the varying cluster-level power requirements across different AI chip and cluster architectures.

To check for consistency between reported and estimated power values, we plotted the correlation below (Figure \ref{fig:reported_vs_calculated_power_capacity}). The correlation coefficient of 0.97 indicates our values are highly correlated.
\vspace{3mm}
{\setlength\intextsep{0pt}
\begin{figure}[H]
    \centering
    \includegraphics[width=0.6\linewidth]{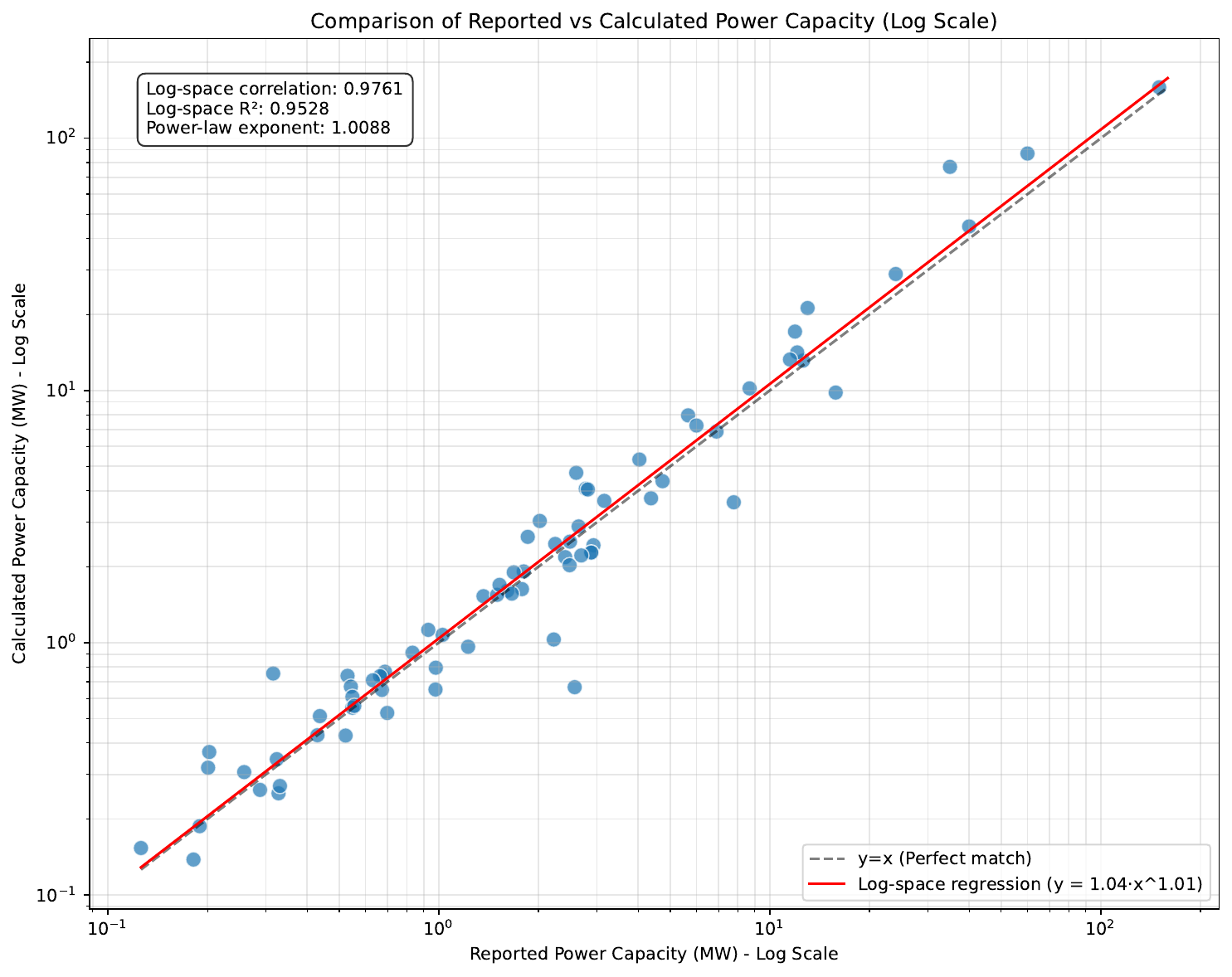}
    \caption{Comparison of power requirements for AI supercomputers that reported it, versus our calculations of power requirements based on chip type and count. }
    \label{fig:reported_vs_calculated_power_capacity}
\end{figure}
}

Note that our methods assess theoretical peak power usage when all the processors are fully utilized and not power consumption. The average power consumption of an AI supercomputer is usually only a fraction of its peak.

\subsection{Hardware cost}
\label{sec:Appendix:Methods:Cost}

We use the publicly reported total hardware cost of the AI supercomputer in our analysis whenever it is available. When it is unavailable, we estimate this cost based on the chip type, quantity, and public chip prices. The procedure used to estimate costs is adapted from \citet{cottier2024historical}. Using Epoch’s \href{https://airtable.com/appDFXXgaG1xLtXGL/shrBLvaZ4E04ao4oo}{dataset of hardware prices}, we select the latest known price of the chips used in the AI supercomputer, from before the system’s first operational date. For each type of chip, we multiply the cost per chip by the number of chips, multiply by factors for intra-server and inter-server overhead, and then sum these costs if there are multiple types of chips. Intra-server cost overhead was estimated in \citet{cottier2024historical} for the NVIDIA P100 (1.54$\times$), V100 (1.69$\times$), and A100 (1.66$\times$), based on known DGX and single-GPU prices near release. We use the mean of these factors (1.64$\times$) for all chips, to estimate server prices, including interconnect switches and transceivers. Then, we adjust for the cost of server-to-server networking equipment, which was estimated to be 19\% of final hardware acquisition costs.

Additionally, we apply a discount factor of 15\% to the final hardware cost of the AI supercomputer to account for large purchasers of AI chips often negotiating a discount on their order. We discuss limitations with this estimate and our cost data in the next section.

Our final formula for estimating hardware cost is as follows:
\vspace{3mm}
\setlength{\abovedisplayskip}{0pt}
\setlength{\belowdisplayskip}{0pt}
\begin{align*}
\textit{Hardware acquisition cost} = & \left[ (\textit{Primary AI chip cost} \times \textit{Primary AI chip quantity}) \right.\\
& \left.+ (\textit{Secondary AI chip cost} \times \textit{Secondary AI chip quantity}) \right] \\
& \times \textit{Intra-server overhead} \times \textit{Inter-server overhead} \times \textit{Discount factor}
\end{align*}

In this formula, our intra-server overhead, or “chip-to-server” factor, is 1.64×, our inter-server overhead, or “server-to-cluster” factor, is 1.23×, and our discount factor is 0.85×.

Notably, our cost figures refer only to the hardware acquisition cost of the AI supercomputer, and not costs required for maintenance, electricity, or the cost of the datacenter hosting it.\footnote{A 2025 estimate of the cost of datacenters puts them at \$11.7 million per MW. This could be combined with our power requirement estimates to get an estimate of hardware plus datacenter acquisition \citep{mcwilliams2025cg}.}

All cost values are adjusted for inflation into 2025 USD, using the producer price index for the Data Processing, Hosting, and Related Services industry, reported by the Federal Reserve Bank of St. Louis \citep{usbls2025}. We divided pre-2025 cost figures by the price index value at its closest reported date and multiplied by the price index value in January 2025. Our trends and forecasts refer to values in 2025 USD.

\subsubsection{Limitations with our hardware cost data}
\label{sec:Appendix:Methods:Cost:Limitations}

Our cost data for AI supercomputers has several important limitations:

1. We found reported cost figures for only a limited subset of AI supercomputers, with data predominantly from public sector systems rather than industry deployments.

2. The reported figures may diverge from true costs in multiple ways. 
    \begin{itemize}[topsep=0pt]
        \item[{\textbullet}]They sometimes represent planned contract costs rather than final realized expenditures. 
        \item[{\textbullet}]Contract figures may bundle additional expenses, such as multi-year operational costs, that should be excluded from our analysis. 
        \item[{\textbullet}]When uncertainty about the precise meaning of reported costs is too high, we excluded the data, though some ambiguity likely remains.
    \end{itemize}
3. We also encountered challenges with estimating hardware costs based on chip quantities and prices. 
    \begin{itemize}[topsep=0pt]
        \item[{\textbullet}]Our price dataset lacks information for some GPU types, particularly custom silicon, though it does cover most common GPUs.
        \item[{\textbullet}]Google does not sell TPUs, so our price data for them is based on comparison of their performance and manufacturing costs with those of NVIDIA chips that have similar technical specifications.
        \item[{\textbullet}]Most GPU suppliers do not publish wholesale prices, forcing us to rely on third-party retailer prices and reports from experts that can vary significantly by vendor and time. 
        \item[{\textbullet}]We use the most recent listed price for each GPU, but prices fluctuate substantially with market conditions, so our limited time-series data means some AI supercomputer costs may be mismatched with the prices actually paid for the chips.
    \end{itemize}
4. Given limited data, we assume that all AI supercomputers have the same overhead costs, but this is unlikely, particularly for systems built five years ago.

5. The discount factor is another significant source of uncertainty. Price negotiations generally occur privately, making reliable estimates difficult, and discounts vary substantially by supplier, purchaser, chip type, and time. For simplicity and due to data limitations, we apply a constant 15\% discount rate across all AI supercomputers, but we expect the true rate to vary significantly by AI supercomputer. We selected this rate because it best aligns with the difference between our cost estimates and reported costs, and stated estimates of discount rates.\footnote{Citi Analysts imply that Microsoft received a 33\% discount compared to other purchasers, who paid what we would count as the full price \citep{tomshardwareNvidiasH100}. If these groups buy equal amounts of chips, this implies an average discount of 16\% \citep{nextplatformStackingMI200}.} However, as stated above, our reported cost data is itself biased. Our universal discount rate likely overestimates costs for major purchasers like U.S. national labs\footnote{NextPlatform implies that the Oak Ridge National Lab Summit supercomputer got close to a 50\% discount on the cost of their GPUs, and that industry partners have historically paid \citep{historicallypaid} 1.5$\times$ to 2$\times$ more for chips than National Labs.} and the largest GPU buyers while underestimating costs in other scenarios.

As a consequence of these limitations, we estimate that a 90\% confidence interval for the true hardware cost value is +/- 0.5 orders of magnitude (within a factor of $\sim3\times$) of our estimate.

\subsection{Forecasts}
\label{sec:Appendix:Methods:Extrapolations}

We extrapolate our observed trends by using the leading AI supercomputer as of March 2025 (xAI’s Colossus) and assuming trends continue until 2030. E.g., for the number of chips, we assume 200,000 chips in March 2025 and multiply this number by 1.6$\times$ each year. Note our approach to extrapolations is simplistic and can only provide rough estimates for future values.

\subsection{Figures and regressions}
\label{sec:Appendix:Methods:Creating_figures}

For all figures and regressions, we filtered the dataset as follows:
\begin{enumerate}[topsep=0pt,itemsep=0pt]
    \item We excluded 99 AI supercomputers where the "Exclude" field is marked. 85 of these systems are outside of our definition because they do not meet our performance threshold. We also excluded 14 systems for other reasons, such as because we decided the chips they used did not qualify as AI chips.
    \item We further excluded 92 AI supercomputers marked as "Possible duplicates". (We try to only mark systems as potential duplicates if we think there is a $>$25\% chance they are a duplicate.)
    \item We further excluded 36 AI supercomputers where "Single cluster" is marked as "No" or "Unclear". 
    \item We excluded 15 AI supercomputers where "Certainty" is lower than "Likely".
    \item We excluded 113 AI supercomputers where "Status" is "Planned", i.e., systems that were not yet operational as of March 2025.
\end{enumerate}

In total, we include 470 out of the 825 systems in our dataset in the analysis. Of these, 389 became operational in 2019 and after.

For all regressions, we consider the 57 AI supercomputers that were in the top-10 by 16-bit FLOP/s and became operational between 1-1-2019 and 1-3-2025.\footnote{In some figures we specify that we are showing trends for the 59 AI supercomputers that were in the top-10 considering highest performance across 32, 16, and 8-bit precisions.}

For our distribution figures we consider all 470 systems remaining after filtering, including those that became operational before 2019. We exclude AI supercomputers that were superseded by newer entries after the newer entry's first operational date.

\subsection{Adequately representing performance gains from using lower precision units}
\label{sec:Appendix:Methods:Numerical_precision}

Values in calculations for AI training (such as model weights, gradients, and updates) can be represented in different precisions. This is analogous to how you may represent the same number as “\$15,228,349,053.84” or “\$15 billion”, depending on the context. In this example, the first representation has a much higher precision than the second, but it also takes more memory to store.

Until the 2010s, AI training primarily used relatively high-precision 32-bit number formats but moved to 16-bit representation in the late 2010s\footnote{\citet{micikevicius2017mixed} is an early example of mixed-precision training which moved the most computationally expensive operations to 16-bit.} and currently seems to be moving to 8-bit, thanks to new hardware supporting these precisions and algorithmic innovations to use the new number formats efficiently \citep{huang20208bit, nvidia2023tensor}. Given that working with values in lower precisions requires less memory and computations, AI chips offer much faster performance for calculations in lower precisions. 

The shift in precision used for training in our study period makes it challenging to adequately display performance trends in our data.
\begin{itemize}[topsep=0pt]
    \item[{\textbullet}]If we showed the highest available performance across these three precisions (Max OP/s)\footnote{OP/s stands for operations per second.} it may seem like AI supercomputers that supported 8-bit precision in the early 2020s were more powerful than they actually were in practice, since 8-bit precision was not widely used to train AI models then.\footnote{Specifically, we are unsure when 8-bit training first became widespread. Developers usually do not report what precisions they use to train their models, making it difficult to assess when newly available formats were widely adopted.} If we used this precision-agnostic trend for our forecasts, we would further imply that shifts to lower precisions will continue, but we cannot make any claims about whether or not that will be the case.\footnote{Specifically, we are unsure when 8-bit training first became widespread. Developers usually do not report what precisions they use to train their models, making it difficult to assess when newly available formats were widely adopted.}
    \item[{\textbullet}]Instead, we limit our analysis to performance in 16-bit precision (16-bit OP/s), which 92\% of the AI supercomputers included in our analysis support.\footnote{For comparison, 96\% of AI supercomputers have a performance for Max OP/s (performance across 32, 16, and 8-bit precisions) The remaining AI supercomputers either lack performance data or we only found a performance for 64-bit precision.} However, we acknowledge that only considering 16-bit performance does not adequately show the performance gains AI companies achieved by moving to lower precision.

\end{itemize}
In practice, we find that trends in a) Max OP/s and b) 16-bit OP/s are mostly consistent. We thus use 16-bit OP/s as the default for our trend analysis and forecasts, but discuss the Max OP/s trend whenever it converges.\footnote{Notationally, we generally refer to 16-bit performance as FLOP/s (instead of "OP/s"), since this is is more common terminology.}

Meanwhile, we decided to use Max OP/s for our inclusion criteria, i.e., to select whether or not a given system has at least 1\% of the performance of the leading operational AI supercomputer.

We include an overview table showing all metrics in each of 16-bit FLOP/s, 8-bit OP/s, and Max OP/s in Appendix \ref{sec:Appendix:Additional_data:Different_precisions}.

\section{Limitations}
\label{sec:Appendix:Limitations}

This section summarizes some overall limitations of our data. We discuss limitations with specific parts of our data in the methods section (Appendix \ref{sec:Appendix:Methods}).

\subsection{Summary of limitations}
\label{sec:Appendix:Limitations:Summary}

\subsubsection{We likely only cover about 10-20\% of all AI supercomputers within our definition}
\label{sec:Appendix:Limitations:Summary:Coverage_percentage}

We use four references to assess our coverage:

\begin{itemize}[topsep=0pt]
    \item[{\textbullet}]\textbf{Coverage by chip production:} Our dataset likely covers 20--37\% of all NVIDIA H100s produced until 2025, about 12\% of all NVIDIA A100s produced, and about 18\% of all AMD MI300X produced. Meanwhile, we estimate we cover less than 4\% of Google’s TPUs and very few custom AI chips designed by AWS, Microsoft, or Meta. We also only cover about 2\% of NVIDIA chips designed to be sold in China (including the A800, H800, and H20). Our average coverage of the six chip types we assessed is 11\%.
    \item[{\textbullet}]\textbf{Coverage by company:} The coverage of different companies varies considerably, from 43\% for Meta and 20\% for Microsoft to 10\% for AWS and 0\% for Apple. The coverage of Chinese companies is particularly poor. Our average coverage of 8 major companies is 15\%.
    \item[{\textbullet}]\textbf{Coverage of total 16-bit FLOP/s in China:} Between end of 2020 and end of 2024 we cover between 10-20\% of total Chinese 16-bit FLOP/s based on an estimate by \citet{idc2025ai}.
    \item[{\textbullet}]\textbf{Coverage of largest training runs:} Our dataset contains a matching AI supercomputer for about half of the largest training runs as of March 2025 reported by \citet{epochai2025notable}. However, we only find official confirmation that the system was used for the specific training run for one-third of all models. Coverage of Chinese training runs is slightly better compared to all training runs.
\end{itemize}

Overall, we estimate we cover between 10 and 20\% of all AI supercomputers as of early 2025. For more details on our coverage, see Appendix \ref{sec:Appendix:Limitations:Comparing_to_public_reports}.

\subsubsection{We lack data for key properties}
\label{sec:Appendix:Limitations:Lacking_data}

\begin{itemize}[topsep=0pt]
    \item[{\textbullet}]We cannot reliably determine when an AI supercomputer was first operational. In most cases, we use the date an AI supercomputer was first reported as existing as the “first operational” date. However, owners may sometimes wait several months before publicly announcing their AI supercomputer, or they may announce a system even if it is not yet available. We expect that most of our “first operational” dates will be a few weeks to a few months later than the real date the AI supercomputer came online.
    \item[{\textbullet}]We sometimes need to make assumptions about basic system facts. For instance, owners sometimes report vague chip quantities such as “EC2 UltraClusters are comprised of more than 4,000 latest NVIDIA A100 Tensor Core GPUs” \citep{AWS2020}, or “With thousands of MI300X GPUs available, clusters of any size can be deployed for reliable, high-performance computing.” \citep{vultr2024}. To include such AI supercomputers, we try to make reasonable estimates of the system's chips and performance and explain our reasoning in the notes field.
    \item[{\textbullet}]Our data is incomplete. Some fields in our dataset are only filled for a fraction of systems, such as reported power requirement, reported hardware cost, and location. However, our data captures key statistics like performance and first operational date for more than 95\% of all AI supercomputers that are included in our dataset.
\end{itemize}

\subsubsection{Key reasons for low coverage}
\label{sec:Appendix:Limitations:Lacking_data:Reasons}

Why do we only cover 10--20\% of all AI supercomputers? The following factors contribute to our low data coverage:
\begin{itemize}[topsep=0pt]
    \item[a)] Companies often choose not to report their AI supercomputers publicly. While companies may benefit from increased public and investor attention when they publish information on large AI supercomputers, they may also prefer to keep this information private to maintain ambiguity about their competitive position.
    \item[b)] Companies may only report their largest AI supercomputers. A large fraction of all chips are sold to hyperscalers that have more limited incentives to publish information about their AI supercomputers. While they may benefit from publishing information about their largest systems, they have no incentives to publish about the number and size of smaller AI supercomputers.
    \item[c)] Even if an owner publishes information about an AI supercomputer, our search methods may not find it, especially if the information is published in a language other than English or Chinese.
    \item[d)] Chinese companies may try to avoid scrutiny from U.S. regulators, both for chips that they legally imported, such as NVIDIA’s A800 and H800, as well as illegally imported chips like NVIDIA’s A100 and H100. Chinese companies may have smuggled more than 100,000 AI chips last year \citep{grunewald2025smuggling}. See Appendix \ref{sec:Appendix:Limitations:Chinese_data:Increased_secrecy} for a longer discussion.
\end{itemize}

\subsection{Detailed limitations}
\label{sec:Appendix:Limitations:Detailed_limitations}

This section discusses some of the limitations of our data and analysis in more detail.

\subsubsection{Defining AI supercomputers is challenging}
\label{sec:Appendix:Limitations:Detailed_limitations:Defining_supercomputers}

Ideally, our dataset would only capture systems that can efficiently run large-scale AI training workloads. However, it is difficult to develop a practical definition that captures only such systems based on limited publicly available data. Additionally, some companies, including Google DeepMind and OpenAI, have used AI chips distributed across multiple data center campuses to train large models \citep{moss2023google,dickson2025distributed}. To adequately include relevant AI supercomputers, we considered the following four definitions:
\begin{itemize}[topsep=0pt]
    \item[a)] AI chips within a single building
    \item[b)] AI chips on a single data center campus
    \item[c)] AI chips within a fixed proximity (e.g., 2 or 5 miles)
    \item[d)] No distance limit; an AI supercomputer is any system capable of training large models.
\end{itemize}

We decided to use definition (b), given the following considerations: The single building (a) may miss cases where well-connected accelerators span multiple buildings on the same campus. A fixed proximity definition (c) is not feasible in practice since we do not know the precise physical location of most of the AI supercomputers in the dataset. Finally, a functional definition (d) is difficult to scope because assessing if a given AI supercomputer meets certain thresholds for performance, connectivity, and integrated operation requires data on network architecture and connections between AI supercomputers that public reports almost never provide. At the same time, we think it is useful to include AI supercomputers that meet the theoretical performance threshold but lack adequate network infrastructure, given it is comparatively easy to retrofit the networking equipment (see Appendix \ref{sec:Appendix:Limitations:Detailed_limitations:Theorhetical_performance}).

We thus adopt the contiguous campus definition (b), where accelerators on a contiguous campus linked by high-bandwidth networks operate as a single AI supercomputer. However, there are two remaining limitations to this definition:
\begin{itemize}
    \item[{\textbullet}]\textbf{Limited data:} Public reports seldom include details on facility boundaries or network topology, making it hard to verify the contiguous nature of a campus.\footnote{We found it particularly challenging to verify this for reports from companies and for AI supercomputers in China.} When we are unsure if a reported system may span several campuses, we mark the field “Single Cluster” as “Unclear” (20 entries). We mark the “Single Cluster” field as “No” if we think the report most likely refers to a decentralized system (8 entries).
    \item[{\textbullet}]\textbf{Decentralized training:} Our dataset currently does not capture the fact that AI developers may use multiple AI supercomputers for a training run. To assess which AI supercomputers may be most suitable for decentralized training, we would need additional information on the network bandwidth between them.
\end{itemize}

\subsubsection{Theoretical performance does not necessarily correspond to usefulness for large-scale training}
\label{sec:Appendix:Limitations:Detailed_limitations:Theorhetical_performance}

\textbf{Systems may lack sufficient networking for efficiently running AI training.} Public performance figures do not guarantee efficient large-scale training. Some AI supercomputers may suffer from inadequate networking, which can reduce utilization and prolong training runs \citep{narayanan2021efficient}. However, systems with inadequate networking infrastructure can easily be upgraded by changing the network fabric, usually at a fraction ($\sim$10--20\%) of the total AI supercomputer cost \citep{leptonai2024llm}.

\textbf{Performance on AI training depends on the software stack.} Our analysis compares theoretical performance across hardware types. In practice, actual performance depends on the software stack and how well the hardware supports it. For instance, despite having a higher theoretical performance, SemiAnalysis assessed that AMD’s MI300X is less useful for large-scale AI training than NVIDIA’s H100 \citep{patel2024benchmark}. This software ecosystem gap becomes especially significant when evaluating AI supercomputers across different hardware platforms, as systems based on Chinese AI chips may not achieve their theoretical potential without the mature software infrastructure that NVIDIA's CUDA provides.

\textbf{Theoretical performance does not fully capture AI inference performance.} Our database focuses on systems suitable for AI training. A system's computation performance is not a good proxy for how well it can run AI inference workloads. NVIDIA's H20, for instance, delivers comparable inference performance to the H100 on certain workloads despite having only 1/7th the raw computational power, due to its high memory bandwidth. We recommend differentiating between FLOP/s (or OP/s for 8-bit and lower) when assessing training capabilities and memory bandwidth in Byte/s when assessing inference and long-context capabilities.

\subsubsection{Limitations with our Chinese data}
\label{sec:Appendix:Limitations:Chinese_data}

Despite involving Chinese speakers in our data collection, we encountered several significant challenges in gathering comprehensive data on Chinese AI supercomputers.
\begin{itemize}[topsep=0pt]
    \item[1.] \textbf{Official announcements often lack key data}, such as information on chip type and quantity. Furthermore, reported performance values often do not include precision.
    \item[2.] \textbf{Sources sometimes report aggregate data for several AI supercomputers.} Computing zones that consist of several separate data center campuses sometimes report total computing capacity at an aggregate level rather than breaking down by individual AI supercomputers.
    \item[3.] \textbf{Different conventions.} Chinese sources sometimes use different metrics and reporting standards than Western conventions, sometimes reporting the number of server racks that we cannot easily convert to chip numbers.
\end{itemize}

While we encounter similar issues for AI supercomputers in other countries, they are particularly common in China. However, we estimate that our database covers 10--20\% of Chinese AI supercomputer performance, which is similar to our coverage estimate for U.S. data (see Appendix \ref{sec:Appendix:Limitations:Comparing_to_public_reports}).

\subsubsection{Chinese owners may have become more secretive about their AI supercomputers, but this has not impacted our data coverage}
\label{sec:Appendix:Limitations:Chinese_data:Increased_secrecy}

In the late 2010s and early 2020s, Chinese supercomputer announcements frequently led to U.S. sanctions, with companies like Sugon, Phytium, and several national supercomputing centers being added to the Entity List due to concerns about military use of these systems \citep{bis2019additions, commerce2021additions}. This is likely what caused China to release less information about its AI supercomputers. In 2022, China stopped submitting any systems to the Top500 list \citep{scmp2022china}.

In October 2022, the U.S. first introduced export controls on AI chips and semiconductor manufacturing equipment with the goal of slowing down Chinese advances in AI \citep{allen2022export}. These export controls were strengthened in October 2023 and December 2024 by fixing loopholes and further restricting Chinese import of chip manufacturing tools \citep{dohmen2023export,allen2024export}. Furthermore, to reduce chip smuggling, the United States introduced the AI Diffusion Framework in early 2025, requiring additional countries to file for a license to import U.S. AI chips \citep{heim2025diffusion}. These actions may have incentivized Chinese owners to further increase secrecy about their AI supercomputers to reduce scrutiny from the United States, particularly if they deployed smuggled AI chips.

However, the effects of increased Chinese secrecy on our data coverage are limited. While we see a decrease in the number of Chinese systems added to our database in 2021 and 2022, the number of Chinese systems increased again in 2024 (Figure \ref{fig:number-of-Chinese-and-US-systems-added-per-year}). Comparing the aggregate performance in our database with \citet{idc2025ai}'s estimate of total 16-bit FLOP/s in China indicates that our coverage was consistently between 10 and 20\% of Chinese performance (see Table \ref{table:chinese_data_compared_to_IDC_estimates}).
\vspace{2mm}
{\setlength\intextsep{0pt}
\begin{figure}[H]
    \centering
    \includegraphics[width=0.8\linewidth]{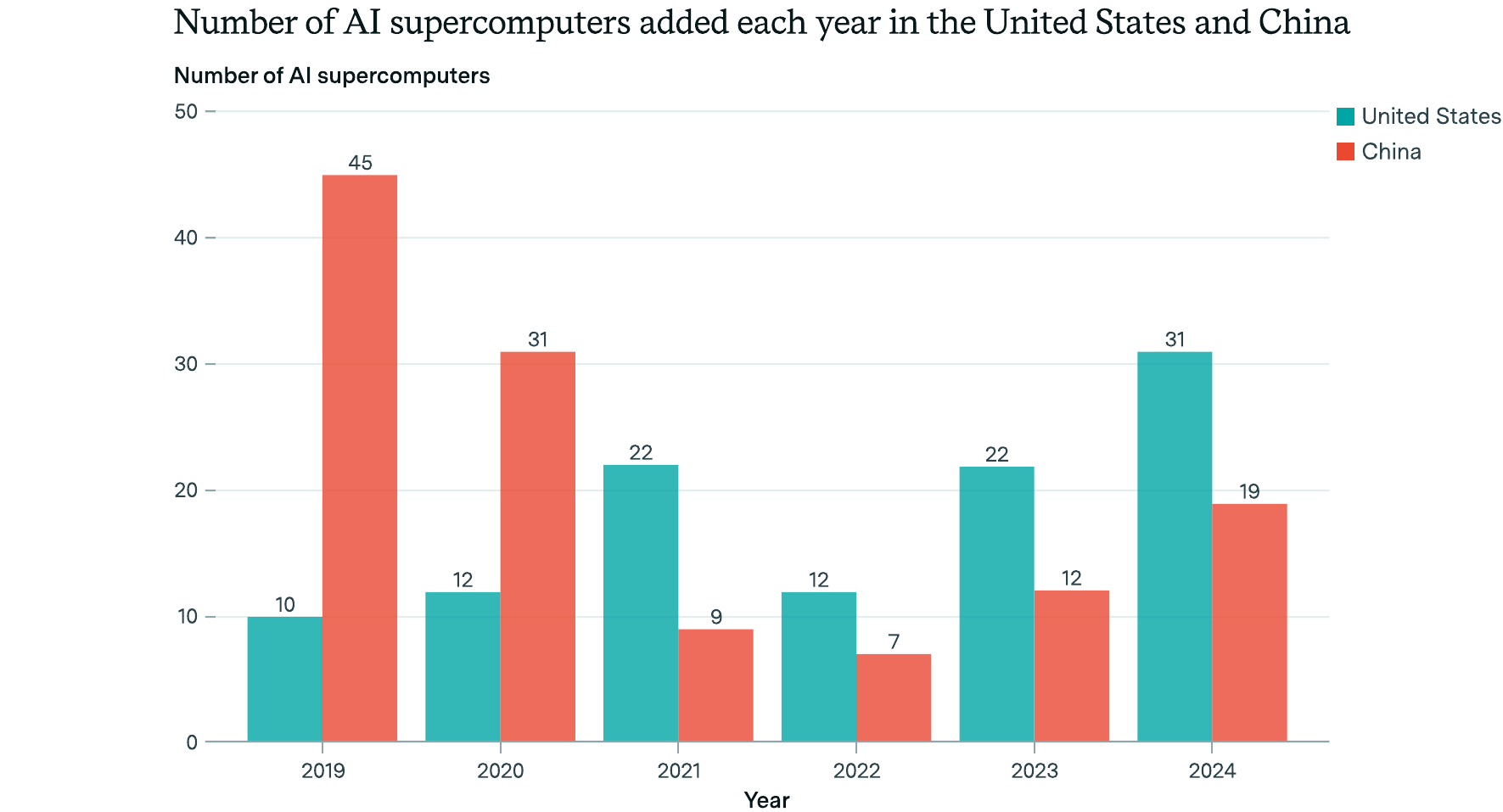}
    \caption{Number of Chinese and U.S. systems added each year.
}
    \label{fig:number-of-Chinese-and-US-systems-added-per-year}
\end{figure}
}

\subsection{Comparing our data with public reports}
\label{sec:Appendix:Limitations:Comparing_to_public_reports}

To assess what fraction of AI supercomputer capacity we capture in the dataset and how our coverage differs between chip types and companies, we compare our data to four sources of public information:
\begin{itemize}[topsep=0pt]
    \item[{\textbullet}]Estimates of the total production of AI chips.
    \item[{\textbullet}]Estimates of the total AI chip stock of companies.
    \item An estimate of the total 16-bit FLOP/s in China by \citet{idc2025ai}.
    \item[{\textbullet}]The fraction of the largest publicly known AI models that were likely trained on an AI supercomputer in our dataset.
\end{itemize}

\subsubsection{Estimating the coverage of all AI supercomputers based on total chip production}
\label{sec:Appendix:Limitations:Comparing_to_public_reports:total_chip_production}

One relevant reference point for our coverage is what fraction of total production we cover for different chip types (Table \ref{table:coverage_by_chip_type}). While some AI chips may be sold to individuals and small research groups, we expect that the vast majority of all AI chips will be used in AI supercomputers that would fall within our definition.

{\setlength\intextsep{0pt}
\begin{table}[H]
\vspace*{3mm}
\caption{Variation of coverage by chip type based on public reports of AI chip production until 2025. Note that the public estimates may include chips in AI supercomputers that are not yet operational or that are otherwise outside of our inclusion criteria. Our full dataset includes potentially existing and planned systems and has a higher coverage. Note that we explicitly search for the H100, A100, and V100 in our automated methodology. This may marginally increase our coverage of these three chip types compared to others.}
\label{sample-table}
\begin{center}
\begin{small}
\begin{sc}
\begin{tabular}{lcccr}
\hline
\abovespace\belowspace
\textbf{Chip type} & \textbf{Public estimate} & \textbf{Dataset} & \textbf{Implied coverage} \\
\hline
\abovespace
H100/H200 & 2.5M -- 4.5M\tablefootnote{Public sources estimate that NVIDIA shipped about 500k H100s in 2023 and 2 million in 2024, for a total of 2.5 million H100s \citep{nolan2023nvidia,shilov2023nvidia}. However, \citet{tweaktown} estimates NVIDIA produced up to 1.5 million H100s in Q4 of 2024. Assuming NVIDIA produced about 1M H100s on average per quarter in 2024 yields a total of 4.5 million H100s. } & 830k & 36.5\% -- 20.3\% \\
A100 & 1.5M -- 3M\tablefootnote{Reports on how many A100s NVIDIA produced are limited, but the company reportedly shipped 500k in Q3 2023 \citep{reportedly}. The A100 was first produced in 2020 and likely reached peak production in 2023 before demand reduced in 2024. It thus seems plausible that NVIDIA produced between 1.5 -- 3 million A100s until 2025.} & 234k & 16.1\% -- 8.1\% \\
H20 & 1M\tablefootnote{\citet{financialtimes2023}} & -- \tablefootnote{We capture 30k H20s that DeepSeek likely owns, but exclude these from the analysis because we are uncertain if they are in the same location.} & 0\% \\
H800/A800 & $>$200k\tablefootnote{Public reports indicate Chinese companies spent \$5 billion on NVIDIA H800 and A800 in 2023 \citep{5billionspent}, indicating at least 200k of these chips imported (conservative estimate assuming \$25k average price per chip \citep{25kavg} .)} & 2k & $<$1.5\% \\
AMD MI300 & 400k\tablefootnote{AMD to ship up to 400,000 new AI GPUs in 2024 \citep{shippinggpu}.} & 72k & 18\% \\
Google TPUs & $>$4M\tablefootnote{Google's internal TPU production likely reached 2 million TPUs in 2023 \citep{martin2024}, although public data is severely limited, given Google does not sell TPUs to outside companies. Assuming a similar production in 2024, there would be at least 4 million TPUs.} & 95k & $<$4\% \\
\belowspace
Other custom silicon & ?\tablefootnote{Microsoft, AWS, and Meta all developed their own custom silicon AI chips deployed in-house \citep{BWPY2024,amazonAcceleratorTrainium,tal2024nextgen}, but we were unable to find trustworthy public estimates of the total numbers.} & 4k & ?\% \\
\hline
\belowspace
\abovespace
Total & 9.6 -- 13.1M & 1.2M & 9.2\% -- 12.5\% \\
\hline
\end{tabular}
\end{sc}
\end{small}
\end{center}
\vskip -0.1in
\label{table:coverage_by_chip_type}
\end{table}
}

Based on the public sources used in the table, our dataset covers between 20\% and 37\% of all NVIDIA H100s produced until late 2024.\footnote{We do not account for H100s produced in 2025, since these would unlikely be installed in any systems before our March 1st cutoff.} However, coverage is much worse for NVIDIA’s H20, A800 and H800, Google’s TPUs, and other custom silicon chips. The average coverage is about 10\%. (Note that the table above only includes confirmed operational AI supercomputers. Our dataset also contains planned AI supercomputers that make up another 920k H100s and 33k MI300X. Some of those may include chips already included in the production volume estimates.)

Table \ref{table:coverage_by_chip_type} reveals that our dataset likely covers H100, A100, and MI300 equally well, whereas coverage of Google’s TPUs and other custom silicon chips is significantly worse. This is expected, given that NVIDIA and AMD sell their chips to a wide range of customers, incentivizing them to report about successful projects to attract more customers. Meanwhile, Google and other hyperscalers only deploy their chips within the company, offering limited incentives to publish more than a few large AI supercomputers.

\subsubsection{Coverage by company}
\label{sec:Appendix:Limitations:Comparing_to_public_reports:By_company}

Another reference point for our coverage is comparing our chip numbers to the publicly reported numbers of chips acquired by different companies (Table \ref{table:coverage_by_company}). 
We expect that hyperscalers deploy most of their AI chips in AI supercomputers covered by our definition, since even when primarily running inference workloads, they usually deploy thousands of AI chips in the same data center. (Note that the March 2025 inclusion threshold was at 2,000 H100-equivalents but was below 1,000 H100-equivalents until August 2024.)

{\setlength\intextsep{0pt}
\begin{table}[H]
\vspace*{3mm}
\caption{Public reports of number of chips owned for various companies at the end of 2024 and comparison with our dataset\tablefootnote{Note we only include systems in our analysis if we are confident they exist in a single site rather than a distributed system. E.g., AWS announced 20k H100 clusters in 2023, but did not explicitly say whether or not those were on the same data center campus.}}
\label{sample-table}
\begin{center}
\begin{small}
\begin{sc}
\begin{tabular}{lcccr}
\hline
\abovespace\belowspace
\textbf{Company} & \textbf{Public claim} & \textbf{Our dataset} & \textbf{Implied coverage} \\
\hline
\abovespace
Meta & \href{https://www.pcmag.com/news/zuckerbergs-meta-is-spending-billions-to-buy-350000-nvidia-h100-gpus}{350k} H100s & 149k & 42.8\% \\
Microsoft & 475k -- 855k H100\tablefootnote{Microsoft likely made up 19\% of total 2023 NVIDIA revenue \citep{fox2024}. We assume they maintained a 19\% share of revenue throughout 2024, and bought a mix of NVIDIA data center products that is approximately equal to NVIDIA's sales mix. Based on estimates for H100 shipments in our previous section, this indicates Microsoft owns between 475k and 855k H100s.} & 118k & 14\% -- 25\% \\
AWS & \href{https://www.techpowerup.com/330027/microsoft-acquired-nearly-500-000-nvidia-hopper-gpus-this-year#g330027-2}{200k} H100s (in 2024) & -- & 0\% \\
Google & \href{https://www.techpowerup.com/330027/microsoft-acquired-nearly-500-000-nvidia-hopper-gpus-this-year#g330027-2}{170k} H100s (in 2024) & 8k & 4.7\% \\
Apple & \href{https://wccftech.com/apple-ai-servers-for-generative-ai-training/}{180k} H100s\tablefootnote{$\sim$2,500 servers in 2023 and 20,000 servers in 2024 * 8 GPUs per server = 180k.} & -- & 0\% \\
CoreWeave & 175k GPUs\tablefootnote{Estimate, given the claim that most of the 250,000 total GPUs said to be H100s and some H200s \citep{Morgan2025}.} & 57k & 22.8\% \\
ByteDance & 310k Hoppers\tablefootnote{About 50k H100 in 2023 and 240k in 2024 \citep{Pires2023, AleksandarK2024}.} & 8k & 3\% \\
\belowspace
Tencent & \href{https://www.techpowerup.com/330027/microsoft-acquired-nearly-500-000-nvidia-hopper-gpus-this-year#g330027-2}{230k} Hoppers (in 2024) & --\tablefootnote{We identified two Tencent AI supercomputers but were unable to identify the performance or hardware used.} & 0\% \\
\hline
\abovespace
\belowspace
Total & 2.09M -- 2.47M & 0.34 M & 13.8 -- 16.3\%\\
\hline
\end{tabular}
\end{sc}
\end{small}
\end{center}
\vskip -0.1in
\label{table:coverage_by_company}
\vspace{1ex}
{\raggedright \textbf{Note: Public estimates cannot be verified and only serve as an approximate assessment of coverage. Some sources are inconsistent with others.} \par}
\end{table}
}

Table \ref{table:coverage_by_company} shows that our coverage differs considerably between companies. While we cover almost half of Meta’s H100s, we cover only 5\% of Google’s and none of Apple’s H100s. Our data is particularly limited for Chinese hyperscalers. However, Table \ref{table:coverage_by_company} does not consider AI supercomputers we cover based on reported performance, but for which we lack the specific chip type. This is especially common for Chinese systems.
\subsubsection{Coverage of Chinese data}
\label{sec:Appendix:Limitations:Comparing_to_public_reports:Chinese_data}

To assess data coverage of AI supercomputers in China, we compare the aggregate 16-bit performance of all Chinese systems in our database to the total Chinese 16-bit performance published in a 2025 report by market intelligence firm International Data Corporation \citep{idc2025ai}. We find that we cover between 10 and 20\% of Chinese 16-bit performance between the end of 2020 and the end of 2024 (Table \ref{table:chinese_data_compared_to_IDC_estimates}). Not all 16-bit performance would likely fall under the definition of our database, so actual coverage of AI supercomputers is likely somewhat higher.

{\setlength\intextsep{0pt}
\begin{table}[htbp]
\vspace{1ex}
\caption{FP16 Performance}
\vspace*{-3mm}
\begin{center}
\begin{small}
\begin{sc}
\begin{tabular}{r c c c}
\hline
\abovespace\belowspace
 & \textbf{Our data} & \textbf{IDC} & \textbf{Implied coverage} \\
\hline
\abovespace
2020 & $1.05 \times 10^{19}$ & $7.50 \times 10^{19}$ & 14\% \\
2021 & $1.88 \times 10^{19}$ & $1.55 \times 10^{20}$ & 12\% \\
2022 & $3.46 \times 10^{19}$ & $2.60 \times 10^{20}$ & 13\% \\
2023 & $4.18 \times 10^{19}$ & $4.17 \times 10^{20}$ & 10\% \\
\belowspace
2024 & $1.46 \times 10^{20}$ & $7.25 \times 10^{20}$ & 20\% \\
\hline
\end{tabular}
\end{sc}
\end{small}
\end{center}
\vspace{1ex}
\label{table:chinese_data_compared_to_IDC_estimates}
\end{table}
}
We were unable to find reliable total performance estimates for other countries, so we had to limit our coverage analysis by FLOP/s to Chinese data.

\subsubsection{Coverage of AI supercomputers used in the largest training runs}
\label{sec:Appendix:Limitations:Comparing_to_public_reports:Training_runs}

To check how well our dataset covers the AI supercomputers used for known large training runs, we check which of the 25 largest training runs in Epoch AI’s \href{https://epoch.ai/data/notable-ai-models}{notable AI models dataset} (as of 1 March 2025) correspond to AI supercomputers in our dataset. (Note that our dataset uses the models dataset as a data source. To avoid circularity we distinguish between systems reported independently from the training run and systems included in our dataset based exclusively on the reports of the training run.)

We find that for about half of the largest AI training runs, we capture an AI supercomputer that could have plausibly been used or was confirmed to be used in the training run (Figure \ref{fig:coverage_of_training_runs_pie_chart}; Table \ref{table:coverage_of_largest_training_runs}).

Our data coverage is slightly better for Chinese AI supercomputers, where we find plausible AI supercomputers for about two thirds of all reported models (Figure \ref{fig:coverage_of_training_runs_pie_chart}; Table \ref{table:coverage_of_largest_chinese_training_runs}).
\vspace{3mm}
{\setlength\intextsep{0pt}
\begin{figure}[H]
    \centering
    \includegraphics[width=0.8\linewidth]{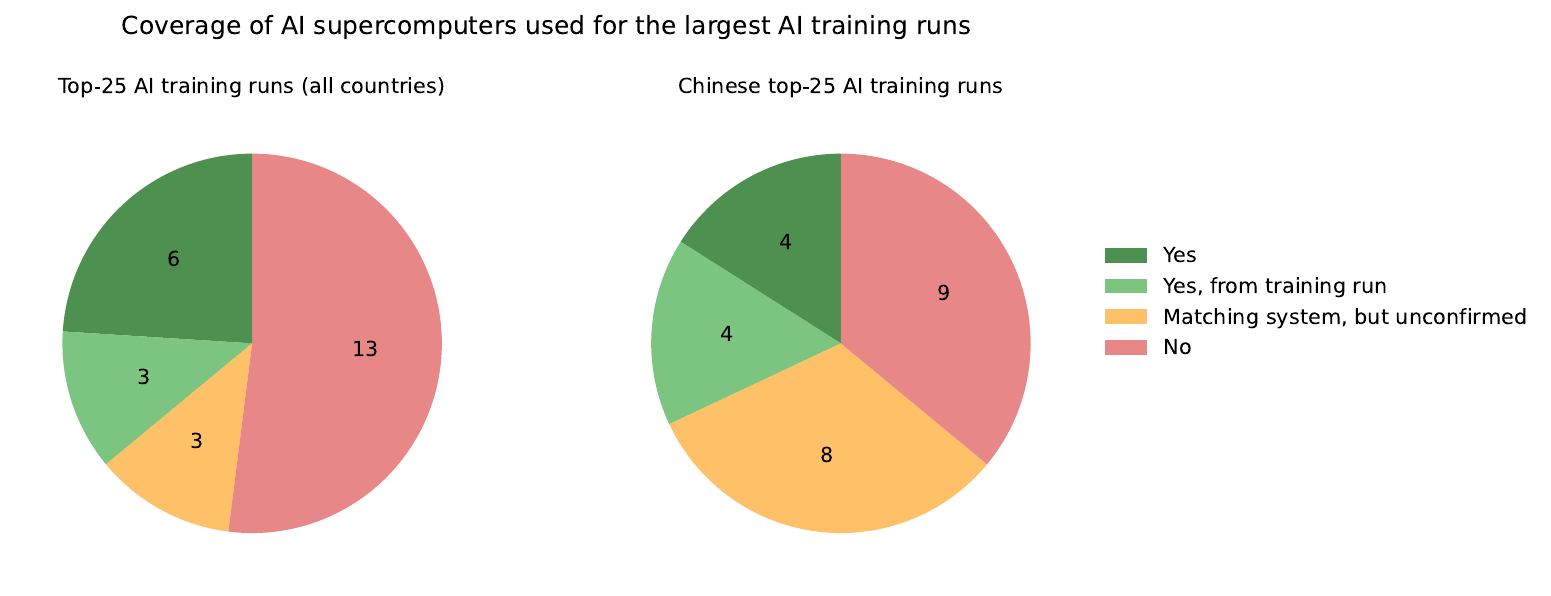}
    \caption{Coverage of AI supercomputers used for the largest AI training runs according to Epoch AI’s notable models dataset. “Yes, from training run” indicates we cover the AI supercomputer, but only based on reports about the training runs itself. “Matching system, but unconfirmed” means an AI supercomputer in our dataset was likely used by the model developer but we find no public reports on whether or not the system was actually used for the training run.}
    \label{fig:coverage_of_training_runs_pie_chart}
\end{figure}
}
\vspace{4mm}
{\setlength\intextsep{0.35pt}
\begin{table}[H]
\caption{Coverage of largest AI training runs (all countries) according to Epoch AI's notable model dataset}
\vspace{-5mm}
\label{sample-table}
\begin{center}
\begin{small}
\begin{sc}
\renewcommand{\arraystretch}{1}  
\setlength{\tabcolsep}{4pt}       
\resizebox{\columnwidth}{!}{
\begin{tabular}{lcr}
\hline
\abovespace\belowspace
\textbf{Training run} & \textbf{Covered} & \textbf{Note} \\
\hline
\abovespace
Grok-3 & \textnormal{Yes} & \textnormal{Trained on Colossus in Memphis, Tennessee} \\
\hline
Gemini 1.0 Ultra & \textnormal{Yes, from training run} & \\
\hline
GPT-4o & \textnormal{No} & \\
\hline
Llama 3.1-405B & \textnormal{Yes} & \textnormal{Presumably trained on Meta GenAI 2024a or 2024b \citep{metagenai2024}} \\
\hline
Claude 3.5 Sonnet & \textnormal{No} & \\
\hline
GLM-4-Plus & \textnormal{No} & \\
\hline
Claude 3.7 Sonnet & \textnormal{No} & \\
\hline
Grok-2 & \textnormal{Matching AI supercomputer,} & \textnormal{Trained on the Oracle Cloud.} \\
& \textnormal{but unconfirmed} & \textnormal{``Oracle OCI Supercluster H100s" matches}\\
& & \textnormal{the description of the training details \citep{archiveXAIsMemphis}}\\
\hline
Doubao-pro & \textnormal{No} & \\
\hline
GPT-4 Turbo & \textnormal{No} & \textnormal{Possibly trained on same AI supercomputer as GPT-4,}\\
& & \textnormal{but no confirmation}\\
\hline
Mistral Large 2 & \textnormal{No} & \\
\hline
GPT-4 & \textnormal{Yes} & \textnormal{Likely trained on Iowa AI supercomputer \citep{iowacompute}.}\\
& & \textnormal{Entered in the dataset as ``Microsoft GPT-4 cluster"} \\
\hline
Nemotron-4 340B & \textnormal{Matching AI supercomputer,} & \textnormal{ ``NVIDIA CoreWeave Eos-DFW"}\\
& \textnormal{but unconfirmed} & \textnormal{appears to match the \href{https://d1qx31qr3h6wln.cloudfront.net/publications/Nemotron_4_340B_8T_0.pdf}{training description}}\\
\hline
Claude 3 Opus & \textnormal{No} & \\
\hline
Gemini 1.5 Pro & \textnormal{No} & \textnormal{We capture several systems from Google,}\\
& & \textnormal{but none were likely used for this model}\\
\hline
GLM-4 (0116) & \textnormal{No} & \\
\hline
Mistral Large & \textnormal{Yes} & \textnormal{Likely used Leonardo} \\
\hline
Aramco Metabrain Al & \textnormal{No} & \\
\hline
Inflection-2 & \textnormal{Yes, from training run} & \\
\hline
Inflection-2.5 & \textnormal{No} & \textnormal{We capture several of Inflection's systems,}\\
& & \textnormal{but none were confirmed}\\
\hline
Reka Core & \textnormal{Yes, from training run} & \\
\hline
Llama 3.1-70B & \textnormal{Yes} & \textnormal{Presumably trained on} \\
& & \textnormal{Meta GenAI 2024a or 2024b \citep{metagenai2024}}\\
\hline
Llama 3-70B & \textnormal{Yes} & \textnormal{Trained on Meta GenAI 2024a or 2024b \citep{metagenai2024}}\\
\hline
Qwen2.5-72B & \textnormal{Matching AI supercomputer,} & \\
& \textnormal{but unconfirmed} & \\
\hline
\belowspace
GPT-4o mini & \textnormal{No} & \\
\hline
\end{tabular}
}
\end{sc}
\end{small}
\end{center}
\vskip -0.1in
\label{table:coverage_of_largest_training_runs}
\vspace{1ex}
\end{table}
}

{\setlength\intextsep{0pt}
\begin{table}[H]
\caption{Coverage of AI supercomputers used for the largest AI training runs in China according to \citet{epochai2025notable} as of March 2025.}
\label{sample-table}
\vskip 0.15in
\begin{center}
\begin{small}
\begin{sc}
\renewcommand{\arraystretch}{1.5}  
\setlength{\tabcolsep}{10pt}       
\begin{tabular}{lcr}
\hline
\abovespace\belowspace
\textbf{Model} & \textbf{Covered} \\
\hline
\abovespace
GLM-4-Plus & \textnormal{No} \\
\hline
Doubao-pro & \textnormal{No} \\
\hline
GLM-4 (0116) & \textnormal{No} \\
\hline
Qwen2.5-72B & \textnormal{Matching AI supercomputer, but unconfirmed} \\
\hline
Telechat2-115B & \textnormal{Matching AI supercomputer, but unconfirmed} \\
\hline
DeepSeek-V3 & \textnormal{Yes} \\
\hline
DeepSeek-R1 & \textnormal{Yes} \\
\hline
MegaScale (Production) & \textnormal{Yes, from training run} \\
\hline
SenseChat & \textnormal{Yes} \\
\hline
Qwen2.5-32B & \textnormal{Matching AI supercomputer, but unconfirmed} \\
\hline
Hunyuan-Large & \textnormal{No} \\
\hline
Qwen2-72B & \textnormal{Matching AI supercomputer, but unconfirmed} \\
\hline
Yi-Large & \textnormal{No} \\
\hline
DeepSeek-V2.5 & \textnormal{Matching AI supercomputer, but unconfirmed} \\
\hline
Yi-Lightning & \textnormal{Yes, from training run} \\
\hline
Qwen1.5-72B & \textnormal{Matching AI supercomputer, but unconfirmed} \\
\hline
Qwen-72B & \textnormal{Matching AI supercomputer, but unconfirmed} \\
\hline
XVERSE-65B-2 & \textnormal{No} \\
\hline
Hunyuan & \textnormal{No} \\
\hline
Luca 2.0 & \textnormal{No} \\
\hline
Qwen2.5-Coder (32B) & \textnormal{Matching AI supercomputer, but unconfirmed} \\
\hline
BlueLM 175B & \textnormal{No} \\
\hline
ERNIE 3.0 Titan & \textnormal{Yes} \\
\hline
MegaScale (530B) & \textnormal{Yes, from training run} \\
\hline
\belowspace
xTrimoPGLM -100B & \textnormal{Yes, from training run} \\
\hline
\end{tabular}
\end{sc}
\end{small}
\end{center}
\vskip -0.1in
\label{table:coverage_of_largest_chinese_training_runs}
\vspace{1ex}
\end{table}
}
\section{Additional data}
\label{sec:Appendix:Additional_data}

\subsection{Overview of trends in different precisions and by sector.}
\label{sec:Appendix:Additional_data:Different_precisions}
\vspace{3mm}
{\setlength\intextsep{0pt}
\begin{table}[H]
\caption{Overview of key trends between 2019 and March 2025. Square brackets indicate the 90\% confidence interval. Note 8-bit trend is only starting in July 2021.\tablefootnote{We assess this trend only after 50 AI supercomputers in our dataset support 8-bit precision.}}
\label{sample-table}
\vskip 0.15in
\begin{center}
\begin{small}
\begin{sc}
\begin{tabular}{lccr}
\hline
\noalign{\vskip 2mm}    
\multicolumn{4}{c}{\textbf{Leading AI supercomputers (including both public and private)}} \\
\noalign{\vskip 2mm}    
\hline
\abovespace\belowspace
 & \textbf{16-bit OP/s} & \textbf{8-bit OP/s} & \textbf{Max OP/s} \\
\hline
\abovespace
Performance Growth & \textbf{2.54} [2.35--2.74] & \textbf{2.60} [2.31--2.93] & \textbf{2.55} [2.34--2.78] \\
Number of Chips & \textbf{1.60} [1.45--1.78] & \textbf{1.69} [1.47--1.94] & \textbf{1.46} [1.29--1.64] \\
Performance per Chip & \textbf{1.60} [1.49--1.71] & \textbf{1.54} [1.42--1.67] & \textbf{1.77} [1.62--1.94] \\
Hardware Cost & \textbf{1.92} [1.76--2.11] & \textbf{1.99} [1.72--2.30] & \textbf{1.76} [1.58--1.97] \\
Cost-Performance & \textbf{1.36} [1.29--1.42] & \textbf{1.37} [1.29--1.45] & \textbf{1.51} [1.43--1.60] \\
Power & \textbf{1.95} [1.77--2.15] & \textbf{2.12} [1.85--2.42] & \textbf{1.78} [1.60--1.99] \\
\belowspace
Energy Efficiency & \textbf{1.34} [1.25--1.43] & \textbf{1.26} [1.20--1.32] & \textbf{1.51} [1.39--1.63] \\
\hline
\noalign{\vskip 2mm}    
\multicolumn{4}{c}{\textbf{Leading private AI supercomputers}} \\
\noalign{\vskip 2mm}    
\hline
\abovespace\belowspace
 & \textbf{16-bit OP/s} & \textbf{8-bit OP/s} & \textbf{Max OP/s} \\
\hline
\abovespace
Performance Growth & \textbf{2.69} [2.47--2.92] & \textbf{3.17} [2.78--3.61] & \textbf{3.00} [2.76--3.27] \\
Number of Chips & \textbf{1.82} [1.66--2.00] & \textbf{2.14} [1.85--2.47] & \textbf{1.83} [1.65--2.03] \\
Performance per Chip & \textbf{1.50} [1.44--1.57] & \textbf{1.48} [1.36--1.61] & \textbf{1.65} [1.55--1.76] \\
Hardware Cost & \textbf{2.06} [1.88--2.26] & \textbf{2.39} [2.09--2.73] & \textbf{2.05} [1.86--2.26] \\
Cost-Performance & \textbf{1.33} [1.28--1.39] & \textbf{1.32} [1.26--1.39] & \textbf{1.47} [1.41--1.54] \\
Power & \textbf{2.16} [1.98--2.35] & \textbf{2.57} [2.26--2.93] & \textbf{2.16} [1.96--2.37] \\
\belowspace
Energy Efficiency & \textbf{1.27} [1.23--1.31] & \textbf{1.23} [1.19--1.28] & \textbf{1.40} [1.34--1.46] \\
\hline
\noalign{\vskip 2mm}    
\multicolumn{4}{c}{\textbf{Leading public AI supercomputers}} \\
\noalign{\vskip 2mm}    
\hline
\abovespace\belowspace
 & \textbf{16-bit OP/s} & \textbf{8-bit OP/s} & \textbf{Max OP/s} \\
\hline
\abovespace
Performance Growth & \textbf{1.86} [1.60--2.15] & \textbf{1.79} [1.46--2.19] & \textbf{1.90} [1.63--2.22] \\
Number of Chips & \textbf{1.21} [0.98--1.50] & \textbf{1.20} [0.96--1.49] & \textbf{1.11} [0.89--1.38] \\
Performance per Chip & \textbf{1.56} [1.34--1.82] & \textbf{1.48} [1.31--1.67] & \textbf{1.75} [1.45--2.11] \\
Hardware Cost & \textbf{1.40} [1.25--1.57] & \textbf{1.34} [1.09--1.65] & \textbf{1.38} [1.20--1.58] \\
Cost-Performance & \textbf{1.41} [1.28--1.56] & \textbf{1.48} [1.32--1.66] & \textbf{1.51} [1.32--1.73] \\
Power & \textbf{1.41} [1.17--1.70] & \textbf{1.38} [1.10--1.74] & \textbf{1.31} [1.07--1.61] \\
\belowspace
Energy Efficiency & \textbf{1.38} [1.19--1.61] & \textbf{1.33} [1.19--1.47] & \textbf{1.56} [1.28--1.90] \\
\hline
\abovespace\belowspace
\end{tabular}
\end{sc}
\end{small}
\end{center}
\vskip -0.1in
\label{table:trends_in_different_numerical_precisions}
\vspace{1ex}
\end{table}
}

\subsection{Chip types in our dataset}
\label{sec:Appendix:Additional_data:Chip_types}

This section covers some additional statistics about the chip types in our dataset. Given the high variance in coverage of different companies and chip types, this data is likely not representative of the broader field.

The majority of chips captured in our dataset are NVIDIA’s Hopper, Ampere and Volta chips (Figure \ref{fig:Distribution_of_chips_in_dataset_pie_chart}; Table \ref{table:chip_type_count_in_dataset}). When grouping similar chips together (such as the H100 and H200), our dataset contains quantities of 27 unique chips. The table below shows all chip types that contribute more than 10,000 chips to our dataset in aggregate. (Note some entries capture chip type but without a known quantity; those are excluded from the table.)

Note that we explicitly search for the H100, A100, and V100 in our automated methodology. This may somewhat increase our coverage of these three chip types compared to others.

{\setlength\intextsep{0pt}
\begin{figure}[H]
    \centering
    \includegraphics[width=1\linewidth]{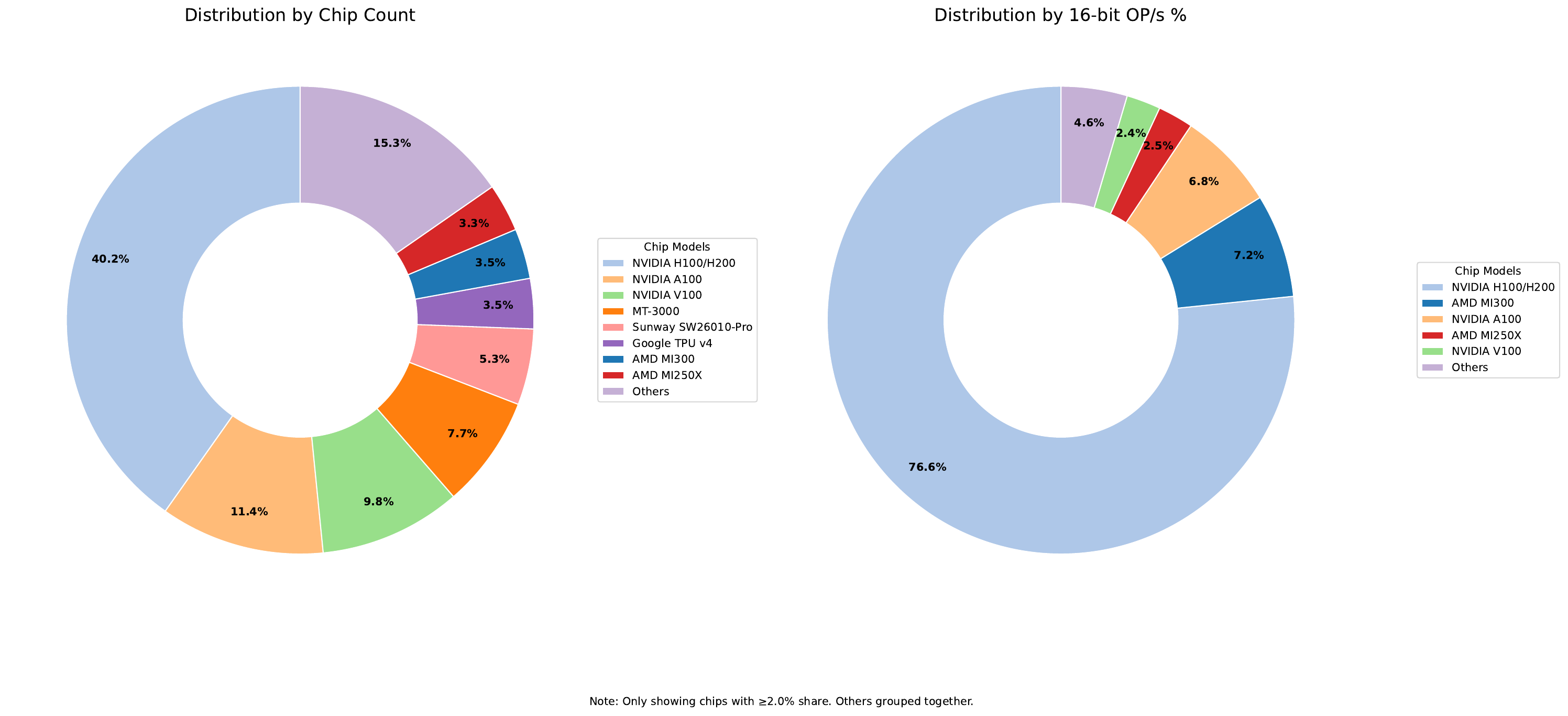}
    \caption{Chip types in operational AI supercomputers in our dataset; by chip count (left) and by FLOP/s (right). Showing only shares above 2\%.}
    \label{fig:Distribution_of_chips_in_dataset_pie_chart}
\end{figure}
}
{\setlength\intextsep{0pt}

\abovespace
\begin{table}[H]
\caption{Table including all chips totaling more than 10,000 in our dataset. Note that this only includes AI supercomputers where chip identity is known. Our dataset also includes AI supercomputers with chip quantities where the chip type is unknown; these account for 4\% of total performance. Performance share is in 16-bit FLOP/s}
\label{table:chip_type_count_in_dataset}
\begin{center}
\begin{small}
\begin{sc}
\begin{tabular}{lccr}
\hline
\abovespace\belowspace
\textbf{Chip Model}          & \textbf{Total number} & \textbf{Performance Share (\%)} \\
\hline
\abovespace
NVIDIA H100/H200             & 830,000  & 76.6 \\
NVIDIA A100                  & 235,000  &  6.8 \\
NVIDIA V100                  & 203,000  &  2.4 \\
MT-3000                      & 160,000  &  0.0 \\
Sunway SW26010‑Pro           & 109,000  &  0.6 \\
Google TPU v4                &  72,000  &  1.8 \\
AMD MI300                    &  72,000  &  7.2 \\
AMD MI250X                   &  69,000  &  2.5 \\
Shensuan‑1                   &  64,000  &  0.2 \\
INTEL MAX 1550               &  64,000  &  0.3 \\
NVIDIA TESLA K20X            &  26,000  &  0.0 \\
GROQCHIP LPU V1              &  20,000  &  0.2 \\
NVIDIA TESLA K40C   & 10,000 & 0.0 \\[0.5ex] \hline
\end{tabular}
\end{sc}
\end{small}
\end{center}
\vskip -0.1in
\vspace{1ex}
\end{table}
}
When sorting chips by manufacturer, we find that NVIDIA chips make up about 75\% of total performance in our dataset (Table \ref{table:chip_count_by_manufacturer_in_dataset}; Figure \ref{fig:distribution_of_chips_by_manufacturer_in_dataset}). This is consistent with NVIDIA’s \href{https://www.cnbc.com/2024/06/02/nvidia-dominates-the-ai-chip-market-but-theres-rising-competition-.html}{2024 AI chip market share} of about 70 - 95\%. Chinese-designed chips make up less than 2\% of the performance in our dataset. However, for Chinese AI supercomputers, we disproportionally lack information about hardware. Thus, a significant share of the unknown chips are likely of Chinese origin.
{\setlength\intextsep{0pt}
\begin{table}[H]
\caption{Table including all chips totaling more than 10,000 in our dataset. Note this only includes AI supercomputers where chip identity is known. The dataset also includes AI supercomputers with chip quantities where the chip type is unknown; these account for 4\% of the dataset by performance. Note that the performance share is in 16-bit performance.}
\label{table:chip_count_by_manufacturer_in_dataset}
\begin{center}
\begin{small}
\begin{sc}
\begin{tabular}{lcr}
\hline
\abovespace\belowspace
\textbf{Manufacturer} & \textbf{Total} & \textbf{Performance Share (\%)} \\
\hline
\abovespace
NVIDIA   & 1,334,962 & 86.1 \\
AMD      &   141,120 &  9.6 \\
SUNWAY   &   108,544 &  0.6 \\
GOOGLE   &    94,856 &  2.3 \\
INTEL    &    67,744 &  0.5 \\
SHENSUAN &    64,000 &  0.2 \\
GROQ     &    19,725 &  0.2 \\
Other    &   215,584 &  0.0 \\[0.5ex]
\hline
\end{tabular}
\end{sc}
\end{small}
\end{center}
\vskip -0.1in
\vspace{1ex}
\end{table}
}
\vspace{8mm}
{\setlength\intextsep{0pt}
\begin{figure}[H]
    \centering
    \includegraphics[width=1\linewidth]{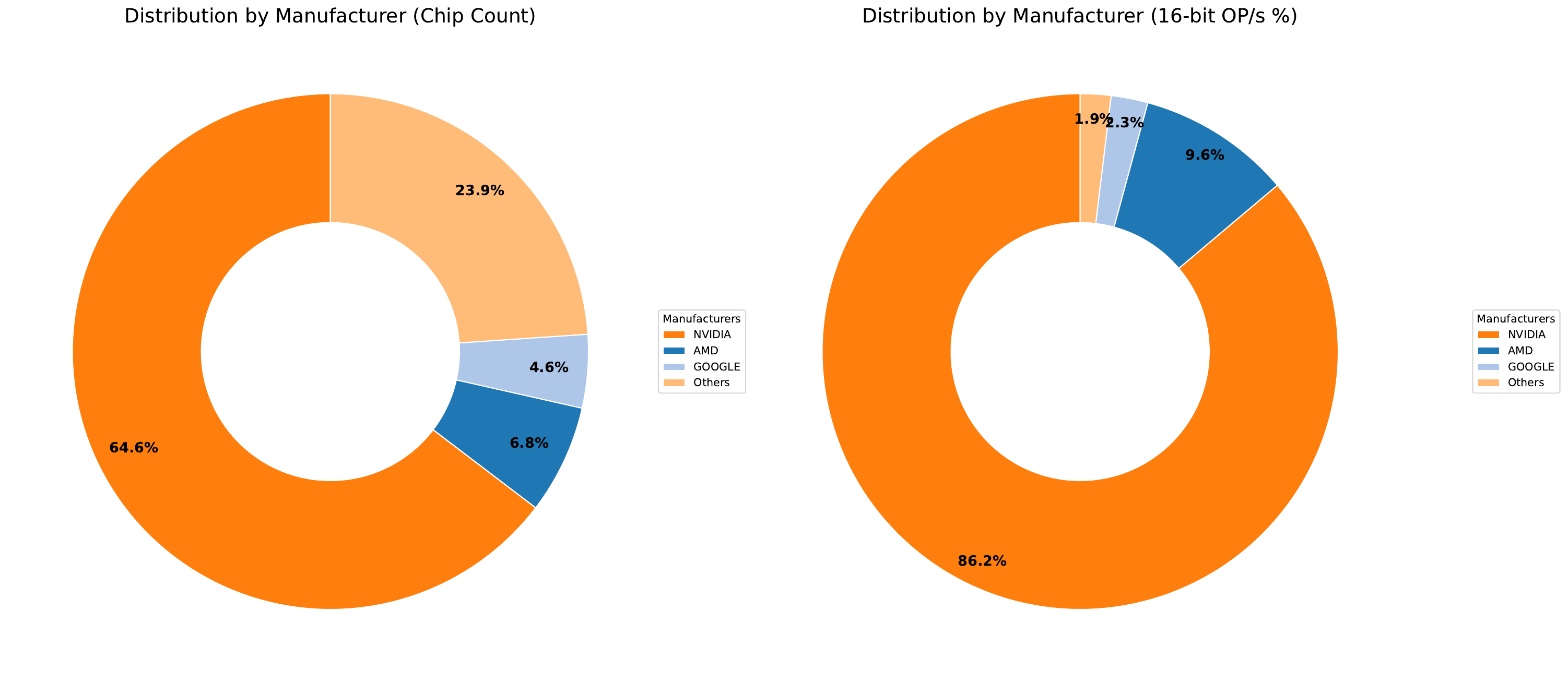}
    \caption{Manufacturer distribution by chip count (left) and by performance (right). We only show manufacturers that contributed at least 2\% of total performance in our dataset.}
    \label{fig:distribution_of_chips_by_manufacturer_in_dataset}
\end{figure}
}
This is consistent with the hardware recorded in \citet{epochai2025notable}, where the vast majority of models with known hardware were trained on NVIDIA chips (Table \ref{table:notable_models_by_hardware_manufacturer}). However, note that the dataset does not focus on hardware, and therefore, the coverage is incomplete.

\begin{table}[H]
\caption{Share of models trained on hardware produced by different manufacturers. Note that hardware information is not recorded for most models.}
\label{sample-table}
\begin{center}
\begin{small}
\begin{sc}
\begin{tabular}{lccr}
\hline
\abovespace\belowspace
\textbf{Manufacturer} & \textbf{Number of models} & \textbf{Share }  \\
\hline
\abovespace
Unknown & 226 & 50.8\% \\
NVIDIA & 144 & 25.6\%  \\
Google & 72 & 16.2\% \\
Huawei & 2 & 0.4\% \\
\belowspace
AMD & 1 & 0.2\% \\
\hline
\end{tabular}
\end{sc}
\end{small}
\end{center}
\vskip -0.1in
\label{table:notable_models_by_hardware_manufacturer}
\vspace{1ex}
\par
\end{table}


\clearpage
\addcontentsline{toc}{section}{References}
\bibliographystyle{icml2024}
\bibliography{refs}
\end{document}